\documentclass[12pt]{article}

\usepackage{txfonts}
\usepackage[]{graphicx}
\usepackage{booktabs}
\usepackage{subcaption}
\usepackage{xcolor}
\usepackage[normalem]{ulem} 
\usepackage{xurl}
 \def\mso{\,\mathrm{M}_\odot}

 \def\kms{\, \mathrm{km}\, {\mathrm s}^{-1}}
 \def\smy{M_\odot\,{\rm yr}^{-1}}
 
 \def\teff{\log\, T_{\rm eff}\,}

  \def\simle{\mathrel{\hbox{\rlap{\hbox{\lower4pt\hbox{$\sim$}}}\hbox{$<$}}}}
 \def\simgr{\mathrel{\hbox{\rlap{\hbox{\lower4pt\hbox{$\sim$}}}\hbox{$>$}}}}

\newcommand*\xbar[1]{%
   \hbox{%
     \vbox{%
       \hrule height 0.5pt 
       \kern0.5ex
       \hbox{%
         \kern-0.1em
         \ensuremath{#1}%
         \kern-0.1em
       }%
     }%
   }%
}

\usepackage{ulem}
\usepackage{color}

\newcommand{\aap}{Astron.~Astrophys.}  
\newcommand{\apjl}{Astrophys.~J.~Lett}   
\newcommand{\apjs}{Astrophys.~J.~Supp.}   

\newenvironment{sciabstract}{%
\begin{quote} \bf}
{\end{quote}}

\title{An X-ray quiet black hole born with a negligible kick
in a massive binary within the LMC}
\author{Tomer Shenar$^\ast$$^{1,2}$,
Hugues Sana$^{1}$,
Laurent Mahy$^{3, 1}$,
Kareem El-Badry$^{4, 5, 6}$,\\
Pablo Marchant$^{1}$,
Norbert Langer$^{7, 8}$,
Calum Hawcroft$^{1}$,
Matthias Fabry$^{1}$,\\
Koushik Sen$^{7, 8}$,
Leonardo A.\ Almeida$^{9, 10}$,
Michael Abdul-Masih$^{11}$,\\
Julia Bodensteiner$^{12}$,
Paul A.\ Crowther$^{13}$,
Mark Gieles$^{14, 15}$,\\
Mariusz Gromadzki$^{16}$,
Vincent H\'enault-Brunet$^{17}$,
Artemio Herrero$^{18, 19}$,\\
Alex de Koter$^{1, 2}$,
Patryk Iwanek$^{16}$,
Szymon Koz{\l}owski$^{16}$, 
Daniel J.\ Lennon$^{18, 19}$,\\
Jes\'us Ma\'iz Apell\'aniz$^{20}$,
Przemys{\l}aw Mr{\'o}z$^{16}$,
Anthony F.\ J.\  Moffat$^{21}$,\\
Annachiara Picco$^{1}$,
Pawe{\l} Pietrukowicz$^{16}$,
Rados{\l}aw Poleski$^{16}$,\\
Krzysztof Rybicki$^{16, 22}$,
Fabian R.\ N.\ Schneider$^{23, 24}$,
Dorota M.\ Skowron$^{16}$,\\
Jan Skowron$^{16}$,
Igor Soszy{\'n}ski$^{16}$,
Micha{\l} K.\ Szyma{\'n}ski$^{16}$,
Silvia Toonen$^{2}$,\\
Andrzej Udalski$^{16}$,
Krzysztof Ulaczyk$^{25}$,
Jorick\ S.\ Vink$^{26}$,
Marcin Wrona$^{16}$\\
\vspace{-0.1cm}
\scriptsize{$^{1}$ Institute of Astronomy, KU Leuven, Celestijnlaan 200D, 3001 Leuven, Belgium}\\
\vspace{-0.1cm}
\scriptsize{$^{2}$ Anton Pannekoek Institute for Astronomy, Science Park 904, 1098 XH, Amsterdam, The Netherlands} \\
\vspace{-0.1cm}
\scriptsize{$^{3}$ Royal Observatory of Belgium, Avenue circulaire/Ringlaan 3, B-1180 Brussels, Belgium} \\
\vspace{-0.1cm}
\scriptsize{$^{4}$ Center for Astrophysics | Harvard \& Smithsonian, 60 Garden Street, Cambridge, MA 02138, USA} \\
\vspace{-0.1cm}
\scriptsize{$^{5}$ Harvard Society of Fellows, 78 Mount Auburn Street, Cambridge, MA 02138} \\
\vspace{-0.1cm}
\scriptsize{$^{6}$ Max-Planck Institute for Astronomy, K\"onigstuhl 17, D-69117 Heidelberg, Germany} \\
\vspace{-0.1cm}
\scriptsize{$^{7}$ Argelander-Institut f\"ur Astronomie, Universit\"at Bonn, Auf dem H\"ugel 71, 53121 Bonn, Germany} \\
\vspace{-0.1cm}
\scriptsize{$^{8}$ Max-Planck-Institut für Radioastronomie, Auf dem Hügel 69, 53121 Bonn, Germany} \\
\vspace{-0.1cm}
\scriptsize{$^{9}$ Escola de Ci\^encias e Tecnologia, Universidade Federal do Rio Grande do Norte, Natal - RN, 59072-970, Brazil} \\
\vspace{-0.1cm}
\scriptsize{$^{10}$ Programa de P\'os-gradua\c{c}\~ao em F\'isica, Universidade do Estado do Rio Grande do Norte, Mossor\'o - RN, 59610-210, Brazil} \\
\vspace{-0.1cm}
\scriptsize{$^{11}$European Southern Observatory, Alonso de Cordova 3107, Vitacura, Casilla 19001, Santiago de Chile, Chile} \\
\vspace{-0.1cm}
\scriptsize{$^{12}$ European Organisation for Astronomical Research in the Southern Hemisphere (ESO), Karl-Schwarzschild-Str. 2,
85748 Garching, Germany} \\
\vspace{-0.1cm}
\scriptsize{$^{13}$ Department of Physics \& Astronomy, Hounsfield Road, University of Sheffield, Sheffield, S3 7RH United Kingdom} \\
\vspace{-0.1cm}
\scriptsize{$^{14}$ ICREA, Pg. Llu\'is Companys 23, E08010 Barcelona, Spain} \\
\vspace{-0.1cm}
\scriptsize{$^{15}$ Institut de Ciencies del Cosmos (ICCUB), Universitat de Barcelona (IEEC-UB), Mart\'i i Franqu\`es 1, E08028 Barcelona, Spain} \\
\vspace{-0.1cm}
\scriptsize{$^{16}$ Astronomical Observatory, University of Warsaw, Al. Ujazdowskie 4, 00-478 Warszawa, Poland} \\
\scriptsize{$^{17}$ Department of Astronomy and Physics, Saint Mary’s University, 923 Robie Street, Halifax, NS B3H 3C3, Canada} \\
\vspace{-0.1cm}
\scriptsize{$^{18}$ Instituto de Astrofísica de Canarias,E-38 200 La Laguna, Tenerife, Spain} \\
\vspace{-0.1cm}
\scriptsize{$^{19}$ Dpto.\ Astrof\'isica, Universidad de La Laguna, E-38\,205 La Laguna, Tenerife, Spain} \\
\vspace{-0.1cm}
\scriptsize{$^{20}$ Centro de Astrobiolog\'ia, CSIC-INTA, Campus ESAC.\ C.\ bajo del castillo s/n, 28692 Villanueva de la Ca\~{n}ada, Madrid, Spain} \\
\vspace{-0.1cm}
\scriptsize{$^{21}$  D\'ept.\ de physique, Univ.\ de Montr\'eal,
QC, H3C 3J7, Canada} \\
\vspace{-0.1cm}
\scriptsize{$^{22}$ Department of Particle Physics and Astrophysics, Weizmann Institute of Science, Rehovot 76100, Israel} \\
\vspace{-0.1cm}
\scriptsize{$^{23}$  Heidelberger Institut f\"ur Theoretische Studien, Schloss-Wolfsbrunnenweg 35, 69118 Heidelberg, Germany} \\
\vspace{-0.1cm}
\scriptsize{$^{24}$  Astronomisches Rechen-Institut, Zentrum f\"ur Astronomie der Universit\"at Heidelberg, M\"onchhofstr. 12-14, 69120 Heidelberg, Germany } \\
\vspace{-0.1cm}
\scriptsize{$^{25}$ Department of Physics, University of Warwick, Gibbet Hill Road, Coventry, CV4 7AL, UK} \\
\vspace{-0.1cm}
\scriptsize{$^{26}$  Armagh Observatory and Planetarium, College Hill, BT61 9DG, Armagh, Northern Ireland
} 
\\
\scriptsize{$^\ast$To whom correspondence should be addressed; E-mail:  T.Shenar@uva.nl}}

\date{\today}

\begin{document} 


\baselineskip24pt


\maketitle 
\bigskip

\begin{sciabstract}

Stellar-mass black holes are the final remnants of stars born with more than 15 solar masses. Billions are expected to reside in the Local Group, yet only few are known, mostly detected through X-rays emitted as they accrete material from a companion star. Here, we report on VFTS 243: a massive X-ray faint binary in the Large Magellanic Cloud. With an orbital period of 10.4-d, it comprises an O-type star of 25 solar masses and an unseen companion of at least nine solar masses. Our spectral analysis excludes a non-degenerate companion at a $5\sigma$ confidence level. The minimum companion mass implies that it is a black hole. No other  X-ray quiet black hole is unambiguously known outside our Galaxy. The (near-)circular orbit and kinematics of VFTS~243 imply that the collapse of the progenitor into a black hole was associated with little or no ejected material or black-hole kick. Identifying such unique binaries   substantially impacts the predicted rates of gravitational-wave detections and properties of core-collapse supernovae across the Cosmos.
\end{sciabstract}

Pairs of stellar-mass black holes in the distant Universe occasionally merge, unleashing bursts of gravitational waves that can be detected here on Earth. 
The number of recorded merger events since their first detection in 2015  is approaching the 100 mark \cite{Abbott2021_third}, and is expected to grow by orders of magnitude in the coming years. In this context, there is an overwhelming international effort aimed at understanding the evolutionary pathways of the merging black holes and the massive stars that formed them \cite{Mandel2022}.
A fundamental uncertainty in this endeavour is whether, and under which conditions, black hole progenitors experience supernova explosions and kicks during core-collapse.  This question has far-reaching consequences: 
from  the observed supernova types and their distributions, through the retainment of black holes in globular clusters, to  the survivability of black-hole binaries in the context of gravitational-wave production and detection rates \cite{Mapelli2022}.
Empirical data on the question of kicks are sparse, largely model dependent, and point in conflicting directions \cite{Israelian1999, Mirabel2003, Gal-Yam2022}. 

Constraints on supernova kicks originate primarily in  X-ray binaries containing a black hole. By construction, black hole X-ray binaries consist of a donor star that transfers mass onto an accreting black hole. In such interacting binaries, strong tidal forces act to circularise the binary orbit and wash away  previous information stored in the eccentricity regarding past kicks. Weakly interacting (and hence X-ray quiescent)  binaries hosting a black hole are therefore indispensable laboratories to tackle the question of kicks empirically.  Such binaries preserve  the black-hole kick signatures in their orbit. Of special interest are massive (O- and early B-type) binaries hosting black holes (OB+BH), which represent a key evolutionary phase towards black-hole mergers \cite{Belczynski2002, Marchant2016}.

The few known X-ray bright OB+BH binaries are thought to constitute the ``tip of the iceberg":  hundreds of X-ray quiet counterparts are predicted to reside in the Milky Way and the Magellanic Clouds \cite{Langer2020}. And yet, we remain blind to this elusive population of binaries.  Only few low-mass binaries with black-hole companions have been reported in the past \cite{Geier2010, Giesers2018, Thompson2019}. In recent years, multiple claims for massive OB+BH binaries in the Milky Way and the Large Magellanic Cloud have emerged \cite{Liu2019, Rivinius2020, Lennon2021, Saracino2022}.  However, virtually all of those reports have been challenged or refuted by follow-up studies \cite{Abdul-Masih2020, Shenar2020LB1,  El-Badry2022Lennon, Bodensteiner2020HR, ElBadry2022NGC1850_BH1}. Aside from a few candidates \cite{Casares2014, Gomez2021} that require confirmation,   massive X-ray quiet OB+BH binaries are not known, let alone outside our Galaxy.

\begin{figure}[!h]
   \centering
\includegraphics[width=\textwidth]{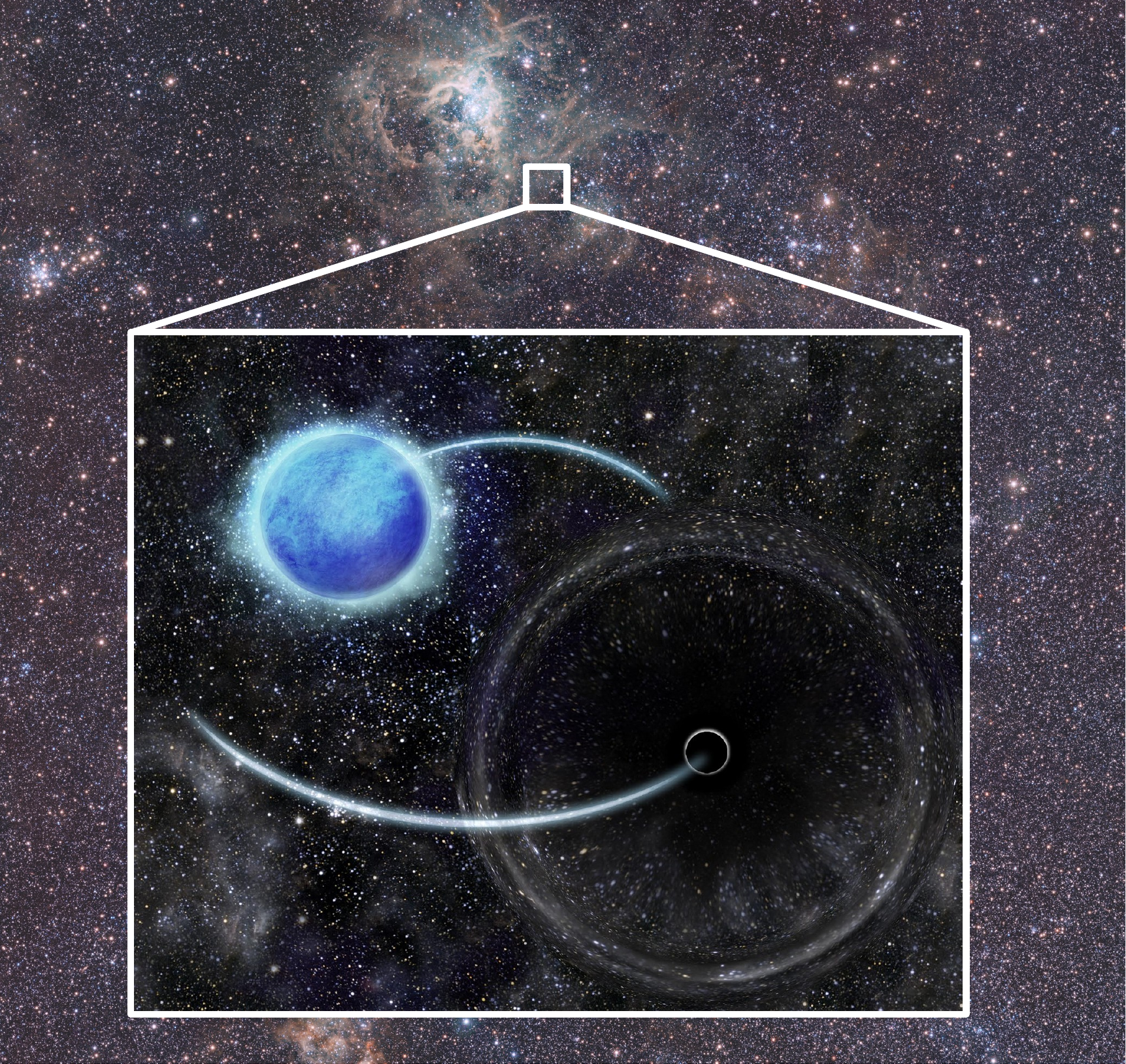}
    \caption{{\bf Visualisation of the O+BH binary VFTS~243.}  The background image shows a Visible and Infrared Survey Telescope for Astronomy (VISTA) image of a segment of the LMC, marking the region in which VFTS~243 resides. The zoom-in shows an artist's impression of the system: a 10.4\,d-period O+BH binary with a (near-)circular orbit. The sizes of the star, black hole, and orbits are not to scale. Background image credit: ESO/M.-R.\ Cioni/VISTA Magellanic Cloud survey. Acknowledgment: Cambridge Astronomical Survey Unit. Visualisation credit: Isca Mayo / Sara Pinilla. 
    } 
    \label{fig:visu}
\end{figure}

Here, we report on the unambiguous discovery of an extragalactic X-ray quiescent O+BH binary, VFTS~243 (Fig.\,\ref{fig:visu}). Located in the Tarantula nebula  in the sub-solar metallicity environment ($Z \approx 0.5\,Z_\odot$) of the Large Magellanic Cloud (LMC), VFTS~243 is one of 51 O-type single-lined spectroscopic binaries (SB1s) characterised by the Tarantula Massive Binary Monitoring (TMBM) \cite{Almeida2017}.
These SB1 binaries  each comprise a well identified massive O-type star orbiting a ``hidden" companion whose spectral signature has not been detected so far.
Using a state-of-the-art analysis method described below, we could unveil the spectral signatures of non-degenerate companions in the vast majority of the 51 SB1 binaries  (Shenar et al.\ in prep.), but not in VFTS~243.  VFTS~243 is the sole target in the sample for which  a black-hole companion provides the only solution consistent with the data. We show this below.

\section*{Results}

We analyse multi-epoch optical spectroscopy acquired with the Fibre Large Array Multi Element Spectrograph (FLAMES) of the European Southern Observatory (ESO) in GIRAFFE mode. Our data cover about six years of observations and consist of five  epochs from the Very Large Telescope Flames Tarantula Survey (VFTS, \cite{Evans2011}) obtained in 2008 and 2009, and an additional 32 epochs from TMBM obtained between 2012 and 2014. The spectra cover the wavelength range 3964--4567\,\AA~at a resolving power of $R = 6400$ and sampling of $\Delta \lambda = 0.2\,$\AA, and have a median signal-to-noise ratio (S/N) of 60 per pixel. We also analyse a light curve obtained by the Optical Gravitational Lensing Experiment (OGLE) \cite{Udalski2015}. 

We derive the orbital elements of the binary (see Supplementary Information) and find results consistent within $1\sigma$ with those previously published \cite{Almeida2017}. VFTS~243 has an orbital period of $P =  10.4031\pm 0.0004\,$d and an eccentricity of $e = 0.017 \pm 0.012$ (errors are 68\%-confidence intervals). Its O7~V primary component displays a radial velocity semi-amplitude of $K_1 = 81.4\pm 1.3\,\kms$, yielding a binary mass function of $f = 0.581 \pm 0.028\,M_\odot$. Analysis of the OGLE light curve further reveals weak ellipsoidal variability with a semi-amplitude of $A_{\rm ellipsoidal} = 0.0015\pm 0.0003\,$mag. 
 
\begin{figure}
   \centering
\begin{tabular}{cc}
\includegraphics[width=0.95\textwidth]{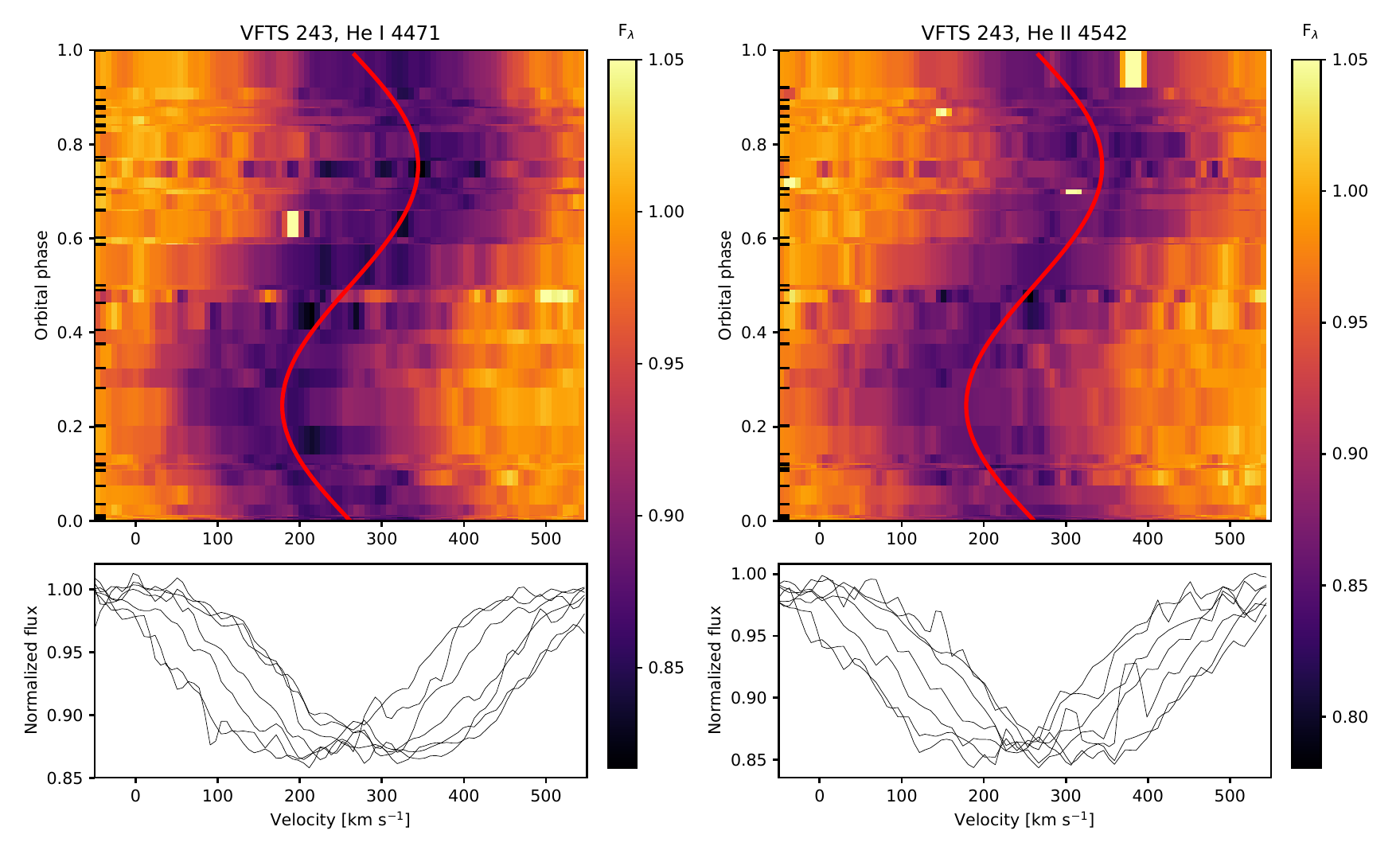}&
\end{tabular}
    \caption{{\bf Dynamical spectra of VFTS~243.} The upper panels show the dynamical spectra of the He\,{\sc i}\,$\lambda 4471$ (left) and He\,{\sc ii}\,$\lambda 4542$ (right) spectral lines in VFTS~243, phased with the system's orbital period of $P=10.4031\,$d. Color-scales indicate the continuum-normalised  flux $F_\lambda$. The red solid line shows the radial-velocity curve of the O7~V primary. The lower panels show individual line profiles at all epochs, binned in phase at $\Delta \phi = 0.1$ for clarity.} \label{fig:VFTS243dyn}
\end{figure}

 While the motion of the O7~V primary is clearly visible in the spectrum, no signature of a secondary object can be immediately identified in the data, confirming its SB1 status (Fig.\,\ref{fig:VFTS243dyn}).  We computed stellar atmosphere models using three independent codes appropriate for O-type stars to constrain the atmospheric properties of the primary star. We derive  the effective temperature $T_{\rm eff, 1} = 36\pm 1\,$kK, surface gravity $\log g_1 = 3.7\pm 0.1$\,[cgs], bolometric luminosity $\log L_1 = 5.20\pm 0.04\,[L_\odot]$, and radius $R_1 =10.3\pm0.3\,R_\odot$.  
 We also detect signatures of CNO-processed material and a  projected rotation velocity of $\varv \sin i = 181\pm 16\,\kms$. The latter value is relatively high compared to single and binary O stars in the Tarantula region.

We estimate the mass of the O7~V component in three different ways (see Methods):  using a spectral-type – mass calibration ($M_{\rm SpT, 1}$),   comparing  $L_1$, $T_{\rm eff, 1}$ to  evolution models ($M_{\rm ev, 1}$), and using the derived surface gravity $\log g_1$ and radius $R_1$  ($M_{\rm spec, 1}$). All estimates yield consistent results within their respective 1$\sigma$ errors (see Table\,\ref{tab:Parameters}). The weighted mean is $M_1 = 25.0\pm 2.3\,M_\odot$. 
The binary mass function and primary  mass yield a minimum mass  of the hidden companion (i.e., at an inclination of $i=90^\circ$) of  $M_{\rm min, 2} = 8.7 \pm 0.5\,M_\odot$, with a 99.7\%\ confidence interval of $7.2 - 10.1\,M_\odot$. By modelling the probability density of $M_1$ as a Gaussian with a mean of $25.0\,M_\odot$ and $\sigma = 2.3\,M_\odot$, and assuming a random orientation of the orbital plane (amounting to a probability distribution of the orbital inclination of $P(i) = \sin i$), we derive the posterior of $M_{2}$ via a Monte Carlo simulation. The posterior has a median and 68\%\ confidence interval of \mbox{$M_{\rm 2} = 10.1_{-2.1}^{+2.0}\,M_\odot$} 
with a mode of $9.0\,M_\odot$. Similarly, the total mass has a median and 68\%\ confidence interval of $M_{\rm tot} = 36.3_{-5.4}^{+3.8}\,M_\odot$ and a mode of $35.0\,M_\odot$.  The faint ellipsoidal variations detected in the light curve of VFTS~243 are consistent with our mass estimates, and imply an orbital inclination in the  range $40-90^\circ$ (see Supplementary Information).

\begin{table}
\centering
\caption{Derived parameters for VFTS~243.  Values related to the primary O7~V component are marked with `1'; those related to the hidden secondary component, with `2'. Errors corresponds to 68\%\ confidence ($1\sigma$) intervals (asymmetric errors are rounded off when differences are small). In addition to the parameters introduced in the text, we also list the argument of periastron ($\omega$), the systemic radial velocity ($\Gamma$),  the colour excess ($E_{B-V}$) and extinction $A_V$, the Roche-lobe filling factor $R_1/R_{\rm Roche~lobe}$, the terminal velocity $\varv_\infty$ and clumping-uncorrected mass-loss rate $\dot{M} \sqrt{D}$ (where $D$ is the clumping factor), and the nitrogen mass fraction $X_N$ (remaining abundances are set to baseline \cite{Brott2011}). For $M_{\rm min, 2}$, $M_2$, and $M_{\rm tot}$, medians are given (modes are in the text).}
\resizebox{.5\textwidth}{!}{\begin{tabular}{lc}\hline \hline
\vspace{-4mm}\\ 
 $P\,$[d] & $10.4031 \pm 0.0004$   \\
$T_0$ [JD-2400000] &  $54870.7 \pm 1.5$   \\
$e$ & $ 0.017 \pm 0.012$   \\
$\omega\,[^\circ]$ & $66 \pm 53$   \\
$K_1\,[\kms]$ & $81.4 \pm 1.3$  \\
$\Gamma \,[\kms]$ & $260.2 \pm 0.9$   \\
$f\,[M_\odot]$ & $0.581 \pm 0.028$   \\ 
$T_{\rm eff, 1}\,$[kK] & $36\pm 1$   \\
$\log g_1\,$[cgs] & $3.7\pm0.1$   \\
$E_{B-V}\,$[mag] & $0.45\pm 0.02$   \\
$A_{V}\,$[mag] & $1.76\pm 0.08$   \\
$\log \dot{M}\,\sqrt{D}$ [$M_\odot\,{\rm yr}^{-1}$] & $-6.3 \pm 0.3$    \\
$\varv_\infty\, [\kms]$ (adopted) & $2100$    \\
$X_{\rm N}$ [\%] &  $0.11 \pm 0.05$    \\
$X_{\rm N}/X_{\rm N, LMC}$  & $16\pm7$   \\
$A_{V}\,$[mag] & $1.70\pm 0.03$   \\
$\log L_1\,[L_\odot]$ & $5.20\pm0.04$   \\
$R_1\,[R_\odot]$ & $10.3\pm0.8$   \\
$R_1/R_{\rm Roche~lobe}$ & $0.33\pm0.03$   \\
$\varv \sin i_1\,[\kms]$ & $181\pm 16$    \\
$M_{\rm SpT, 1}\,[M_\odot]$ & $25.9\pm3.1$   \\
$M_{\rm ev, 1}\,[M_\odot]$ & $26.2\pm2.1$   \\
$M_{\rm spec, 1}\,[M_\odot]$ & $19.3\pm5.2$ \\
$M_{\rm 1}\,[M_\odot]$ (weighted mean) & $25.0\pm2.3$   \\
\vspace{-4mm}\\
$M_{\rm min, 2}\,[M_\odot]$ & $8.7 \pm 0.5$   \\
\vspace{-4mm}\\
$M_{\rm 2}\,[M_\odot]$ & $10.1\pm2.0$   \\
\vspace{-4mm}\\
$M_{\rm tot}\,[M_\odot]$ & $36.3^{+3.8}_{-5.5}$   \\
\vspace{-4mm}\\
$i\,[^\circ]$ & $\gtrsim 40$\\
age\,[Myr] & $7.4$    \\
\hline
\end{tabular}}
\label{tab:Parameters}\end{table}

\begin{figure}
\centering
\includegraphics[width=.7\textwidth]{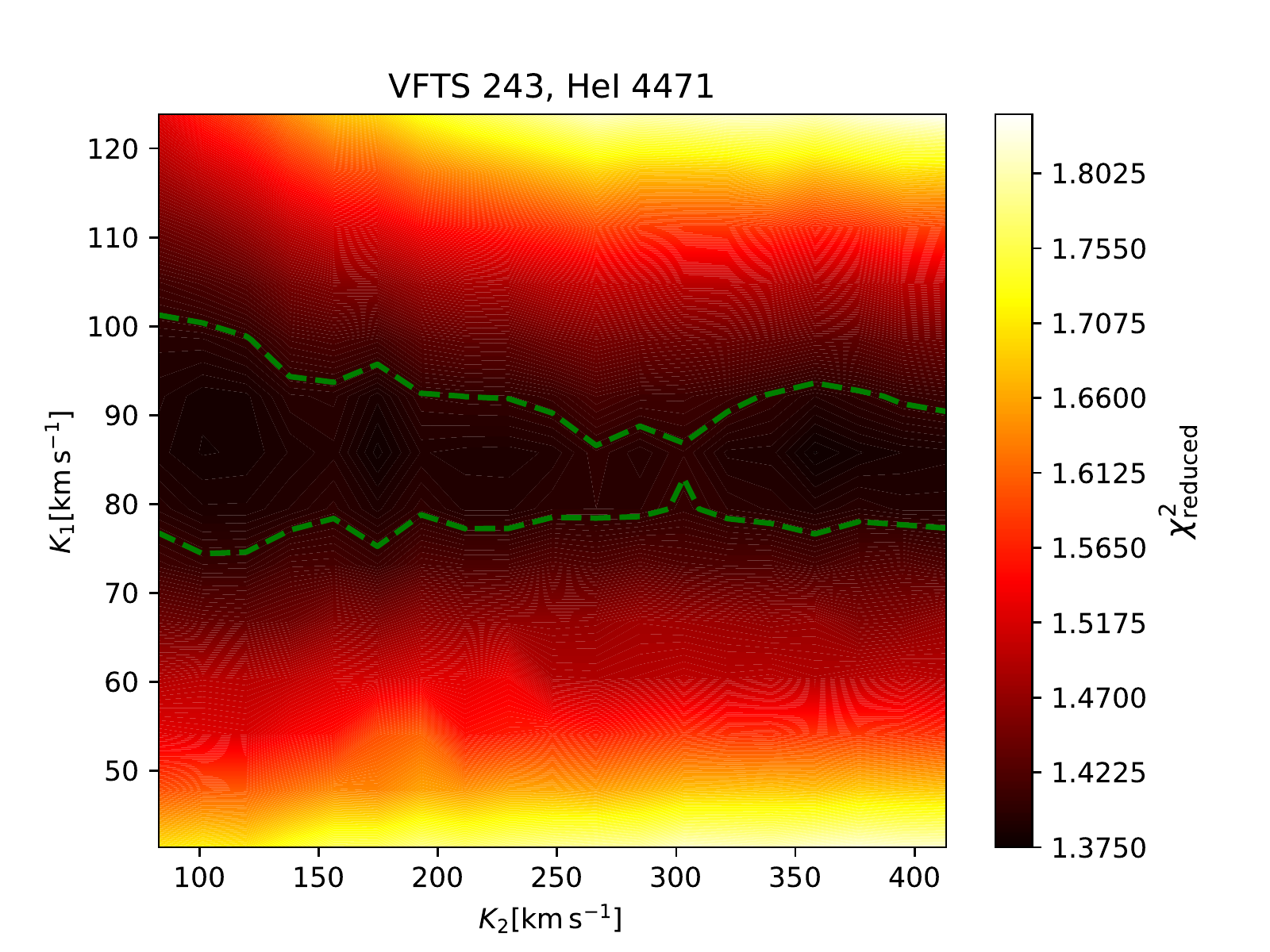}
\caption{{\bf $\chi_{\rm reduced}^2(K_1, K_2)$ map obtained from disentangling the He\,{\sc i}\,$\lambda 4471$ line (Eq.\,\ref{eq:chi2})}. The dashed line shows the 1$\sigma$ contour. $K_1$ is consistent with the value found from the orbital solution, but $K_2$ is not constrained.} \label{fig:chi2243}
\end{figure}

\begin{figure}
\centering
\includegraphics[width=\textwidth]{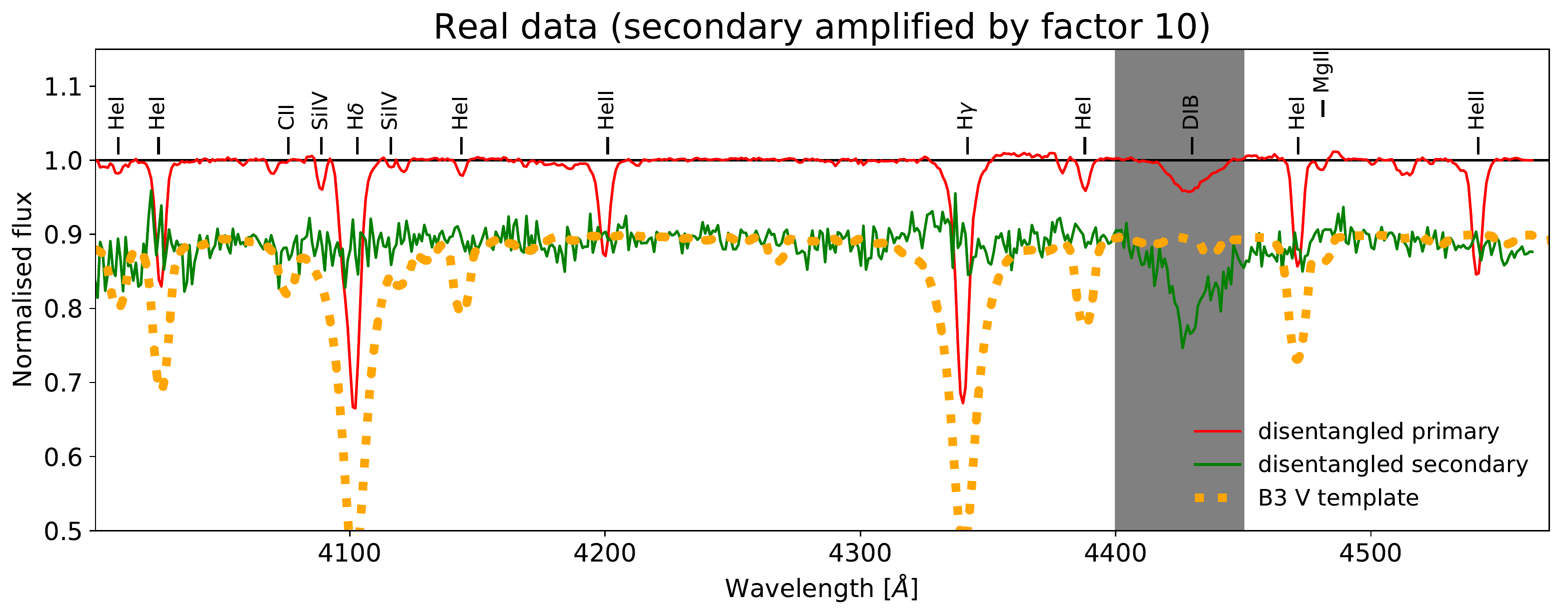}
\caption{{\bf Disentangled spectra of VFTS~243.} The spectra are obtained  for $K_1 = 81.4\kms$ and $K_2 = 3\times K_1 = 244\kms$. The disentangled spectrum of the secondary is amplified by a factor of 10, binned at $\Delta \lambda = 1\,$\AA~and is shifted by $-0.1$ along the vertical axis for clarity. Evidently, it is virtually flat (the one absorption feature seen at $4420\,$\AA~is a diffuse interstellar band). A TLUSTY  template of a $\approx 6.5\,M_\odot$ star (B3~V, $T_{\rm eff} = 19\,$kK, $\log g = 4.0$\,[cgs], $\varv \sin i = 300\,\kms$) is plotted for comparison, showing that such a star cannot be present in the data.  The results are independent of $K_2$. } \label{fig:DisData}
\end{figure}

A main sequence companion accommodating the  constraint on $M_{\rm min, 2}$ would need to have a spectral type earlier than $\approx$ B3~V. More exotically and much less likely than a black hole, the hidden companion may be a massive helium star or a binary in itself.  Should any of these scenarios be viable, the spectral signature of the companion(s) would have been identified by grid spectral disentangling. This technique exploits the Doppler motion of the binary components to extract their individual component spectra \cite{Hadrava1995} (see Methods).
Thanks to the significant enhancement of the S/N that results from combining the entire dataset, the spectral disentangling technique allows us to uncover spectral signatures of objects contributing as little as $\approx 1$\%\ to the optical flux. It has further been successfully applied in the past to unveil the hidden companion stars in other binaries erroneously identified as black hole binaries (e.g., LB-1 \cite{Shenar2020LB1} , HR~6819 \cite{Bodensteiner2020HR}, 2M0412  \cite{El-Badry2022Giraffe}).   Fixing the orbital period and eccentricity, we varied the radial-velocity semi-amplitudes $K_1, K_2$ by minimising the reduced Chi-square statistic $\chi_{\rm reduced}^2(K_1, K_2)$ between the co-added disentangled spectra and each observation. We find that $\chi^2$ is flat with respect to $K_2$, implying that the quality of agreement is independent on $K_2$ (Fig.\,\ref{fig:chi2243}). Regardless of the input value of $K_2$, the extracted spectrum of the secondary is virtually flat, and does not contain features that can be considered of stellar origin (Fig.\,\ref{fig:DisData}).  Hence, the companion is either an optically faint non-degenerate star (or binary) or a compact object.

\begin{figure}
\centering
\includegraphics[width=\textwidth]{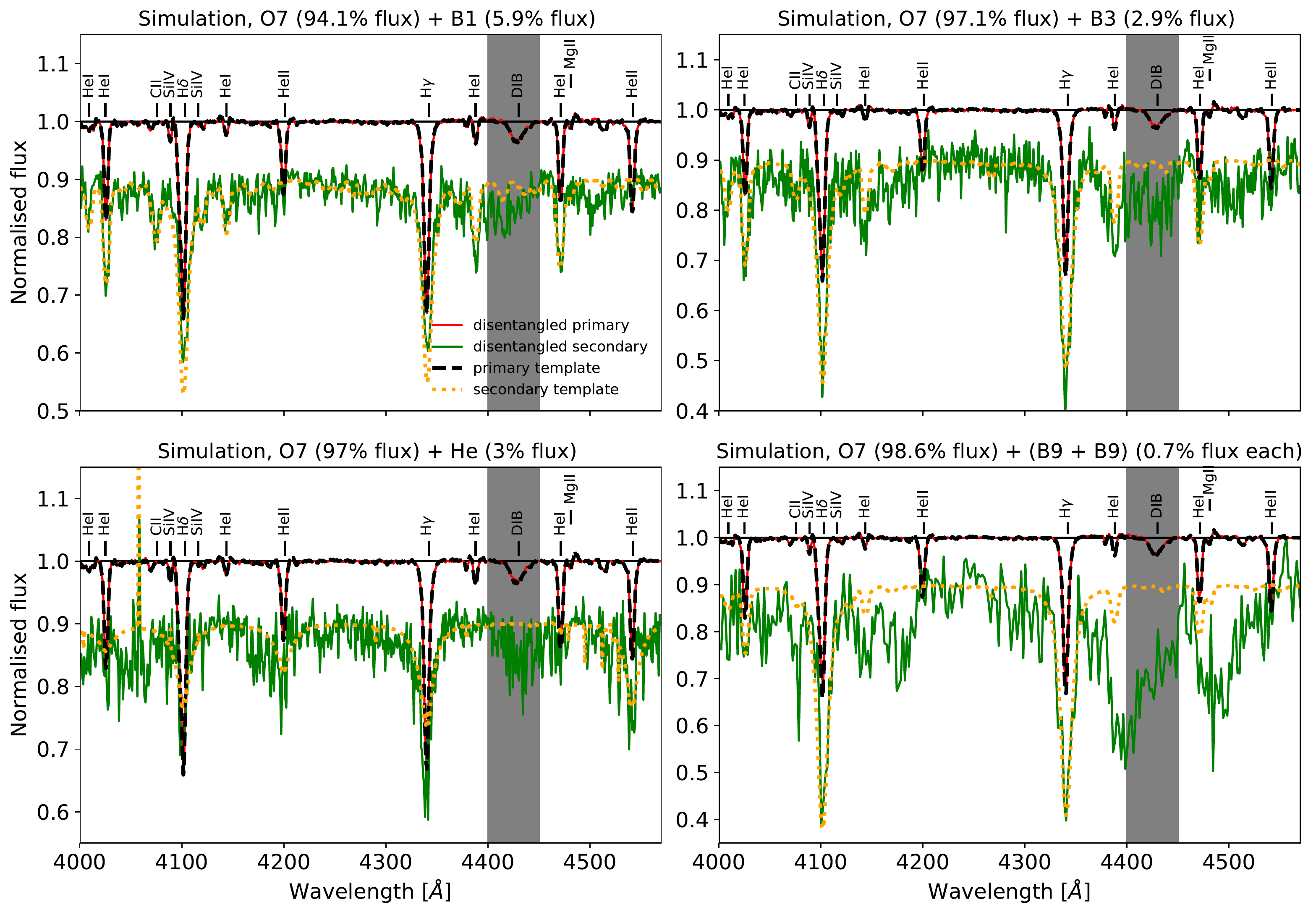}
\caption{{\bf Results from simulated spectra of mock binaries.} Results are shown for:  O7~V + B1~V (upper left, corresponding to $M_2 = M_{\rm min, 2} =  8.7\,M_\odot$), O7~V + B3~V (upper right, corresponding to $M_2 = 6.5\,M_\odot$, our $5\sigma$ lowest limit on the minimum mass), O7~V + helium-star (lower left, $M_{\rm He} = 6.5\,M_\odot$), and the triple scenario O7~V + B9~V + B9~V (lower right, $M_2 = 2\times M_{\rm B9} = 2\times 3.2\,M_\odot$).  The orbital parameters are identical to those of VFTS~243, and the simulated data mimic the real data ( S/N, resolving power) of VFTS~243. The disentangled spectra are compared to the templates used to construct the mock simulations, showing that any plausible companion  abiding to our $5\sigma$ minimum mass constraint of $M_2 > 6.5\,M_\odot$  would be readily retrieved. The spectra are disentangled for $K_2 = 300\,\kms$ (see Supplementary Information), and binned at $\Delta \lambda = 1\,$\AA~for clarity ($\Delta \lambda = 2\,$\AA~for the triple scenario).} \label{fig:Sims}
\end{figure}

To test whether a faint non-degenerate companion could have avoided detection, we produce mock data mimicking the phase coverage and data quality  (i.e., S/N, resolving power) of the real data. We consider  companions  as light as  $M_2 \approx 4\,M_\odot$ (well below our $5\sigma$ threshold), contributing as little as $\approx 1\%$ to the total flux. We  perform simulations by combining spectral templates and synthetic noise, as well as by injecting spectral templates and additional noise into the real data.  
{\it In all simulated cases, the grid disentangling method is able to retrieve the signature of the hypothesised companion without ambiguity, whether it is a main-sequence dwarf, a helium star, or a binary in itself (Fig.\,\ref{fig:Sims}) . This shows that, if it was present in VFTS~243, a non-degenerate companion would have been identified}.
Having rejected all other scenarios,
we are left to conclude that the  hidden companion is a stellar-mass black hole. No other X-ray quiet OB+BH binaries have been unambiguously detected outside our Galaxy so far.

\section*{Discussion}

With over ten refuted claims of X-ray quiescent black-hole binaries in the last two years, one may wonder whether VFTS~243 constitutes yet another false alarm. In this context, it is important to highlight the differences between VFTS~243 and the bulk of previously reported objects.

First, unlike the majority of previously reported X-ray quiescent OB+BH binaries, whose analysis relied on interpretation of complex, near-static  spectral emission lines, VFTS~243  exhibits a regular O-type star spectrum with no remarkable features. The orbital and spectral analysis, the disentangling of the system, and the interpretation of the results are therefore straightforward and leave little room for complex scenarios.

Second, a major culprits of previous erroneous OB+BH claims was that the optically bright primaries in these binaries turned out to be rare, low mass ($M \lesssim 2\,M_\odot$), bloated stars that had been stripped of their outer envelopes \cite{Irrgang2020, Shenar2020LB1}. This led to a severe overestimation of the mass of the bright component, and, in turn, to a false report of black holes. Some of these scenarios further required low orbital inclinations \cite{Liu2019, Lennon2021}, which are less likely from a statistical point of view.  
In contrast, our analysis leaves no room for the primary being a low mass bloated stripped star, and the interpretation of the data does not hinge on an unlikely inclination, but rather points to inclinations larger than $\approx 40^\circ$. In fact, even if the primary mass were to be overestimated, the ellipsoidal variability would require a very massive secondary component ($M_2 \gtrsim 15\,M_\odot)$, which could only be explained if it were a black hole (see Supplementary Information).

Finally, while X-ray quiescent OB+BH binaries are very rare, uncovering them in a sample of 51 SB1 massive binaries is not surprising. Recent predictions suggest that roughly 8\%\ of the 51 SB1 binaries in our sample  should host black holes, amounting to  four expected OB+BH targets in the TMBM SB1 sample \cite{Langer2020}. 

Not only is VFTS~243  an OB+BH binary, but it is also an X-ray faint one. Analysis of deep Chandra observations taken in the context of the Chandra Visionary Program T-ReX
yields an upper limit of $\log L_X< 32.16$\,$[{\rm erg}\,{\rm s}^{-1}$] on the intrinsic X-ray luminosity (Crowther et al.\ 2022, submitted).
Using the derived system parameters and unclumped wind mass loss rate (cf.\ Table\,\ref{tab:Parameters}), we expect a Bondi-Hoyle wind accretion rate of
$1.7 \cdot 10^{-11}\,M_{\odot}\,$yr$^{-1}$ \cite{Bondi1952} .
However, a necessary (yet not sufficient, e.g. \cite{Rodriguez2020}) condition  for producing copious X-rays is the formation of an accretion disk \cite{Shakura1973, Shapiro1986}. Following \cite{Sen2021},  
we find the matter accreted by the BH to have about 30-times too little angular momentum to form an accretion disk. The infalling material is therefore not expected to radiate sufficiently to exceed the detectability threshold (see Supplementary Information).

The relatively rapid rotation and CNO-processed material observed in the O7~V primary suggests that it has previously accreted mass from the black-hole progenitor. Such a  process tends to circularise the binary orbit before the mass-donor undergoes core-collapse.
It is therefore likely that the binary orbit had been circular prior to the core-collapse of the black-hole progenitor. 
In the future, the system will likely  experience a second mass-transfer phase as the O7~V star expands to fill its Roche lobe. The mass ratio of about 2.5 implies that the binary will tighten, and the properties of VFTS~243 suggest that it will end its life as a double black-hole binary within about 5~Myr.  With an expected period of the order of a few days, this newly formed BH+BH binary 
should merge on a timescale of tens to a few hundreds billion years.

The near-circular orbit of VFTS~243 ($e<0.029$; 68\%-confidence interval) has important implications regarding  core-collapse physics of black-hole progenitors. Since the rotation period is not synchronised with the orbit, tidal circularisation can be neglected, implying that the current eccentricity has  not changed since the black hole formed (see Supplementary Information). A mass loss $\Delta M$ during core-collapse  would have induced a change in eccentricity of $\Delta e = e =  \Delta M / M_{\rm tot}$.  Given the current total mass and the  upper-limit on the eccentricity, we derive a maximum mass loss of \mbox{$\Delta M < 1.1\,\,M_\odot$} (68\%-confidence interval). Part of this mass is likely  to have been ejected in the form of neutrino emission, estimated to be of the order of $\Delta M_\nu \approx 0.2-0.5\,M_\odot$ \cite{Lovegrove2013}. Hence, we can conclude that no more than $\approx 0.5\,M_\odot$ of the envelope of the star was removed during the core collapse associated with the formation of the black hole in VFTS~243. 

Small  mass loss and associated kicks were also proposed for the black-hole progenitor in the prototypical Galactic high-mass X-ray binary Cyg~X-1 \cite{Mirabel2003}. However, the latter analysis invokes  a  diverse set of assumptions involving its origin in the Cygnus OB association  and its present-day mass (which has been recently revised \cite{Miller-Jones2021}). Direct core-collapse of black-hole progenitors is currently favoured by   models for stars initially more massive than $\approx 20\,M_\odot$ \cite{Sukhbold2016}.
However, indirect empirical evidence for supernova and/or kicks associated with black holes have also been put forward \cite{Gal-Yam2022, Israelian1999}.  In this context, the properties of VFTS~243 provide direct empirical evidence for the implosion of a massive star into a black hole, and this at sub-solar metallicity. Moreover, the $\approx 10\,M_\odot$ black hole in VFTS~243 probes a mass regime where supernovae may be expected to occur \cite{Sukhbold2016}.

VFTS~243 likely represents a first glimpse of a much larger population of black hole binaries to be detected in the coming years. Upcoming high-precision astrometry of binary systems from the ESA {\it Gaia} satellite  \cite{Gaia2021} is predicted to reveal hundreds of such objects in our Galaxy \cite{Breivik2017, Janssens2022}, while future large-scale spectroscopic surveys (e.g., SDSS/APOGEE, 4MOST), combined with photometric surveys (OGLE, TESS) \cite{Gomel2021} will enable astronomers to filter out additional extragalactic black hole binaries.  These upcoming samples should reveal whether the current properties of VFTS~243 are universally shared among OB+BH binaries. In turn, this will allow astronomers to learn whether core-collapse physics leading to black-hole formation is unique, or whether there is a diversity of pathways, possibly involving parameters such as stellar mass, rotation, and metallicity. Answering these questions is important if we are to properly predict the  properties of newly-formed BH+BH binaries and adequately interpret that vast amount of detections that will be provided by the next generation of gravitational wave observatories.


\newpage

\section*{Methods}

\subsection*{Data}

Our analysis mainly relies on 32 epochs of optical spectra acquired by the TMBM campaign \cite{Almeida2017} between Oct 2012 and Mar 2014 using the fiber-fed multi-object Fibre Large Array Multi Element Spectrograph (FLAMES)  mounted on UT2 of the Very Large Telescope (VLT)   at Paranal observatory, Chile. The spectra cover the wavelength range 3964--4567\,\AA~at a resolving power of $R = 6400$, with a median S/N of 61. Five additional spectra from the VFTS campaign \cite{Evans2011} with similar properties are also used. We extend this range with VFTS
spectra  covering the 4499–5071$\,$\AA  and 6442–6817$\,$\AA~ranges of similar resolution and S/N.  In addition, we make use of a light curve of VFTS~243  obtained from  the Optical Gravitational Lensing Experiment (OGLE) data reductions III and IV \cite{Udalski2015}, as well as a compilation of photometry covering the UV-IR range. More details regarding the acquisition, reduction, and sources of these data are provided in Supplementary Information.

\subsection*{Primary mass}

We provide three mass estimates for the primary star in Table\,\ref{tab:Parameters}. (i) Spectral-type - mass calibration: $M_{\rm SpT, 1}$ is obtained by computing the mean and standard deviation of the evolutionary masses of 19 apparently-single stars in the VFTS sample with similar spectral types (i.e., O6.5~V - O7.5~V) \cite{Schneider2018}.  The masses of these 19 stars were computed  by \cite{Schneider2018} using their measured physical properties (e.g., $L, T_{\rm eff}, g$) as input in the BONNSAI Bayesian tool (\url{www.astro.uni-bonn.de/stars/bonnsai}) \cite{Schneider2014}, which estimates the evolutionary mass using single-star evolution tracks from \cite{Brott2011} and \cite{Koehler2015}.  (ii)  Evolutionary mass: $M_{\rm ev, 1}$ is obtained by inputting our derived values of $\log L_1$ and $T_{\rm eff, 1}$ for VFTS~243 in the BONNSAI tool;  (iii)  Spectroscopic mass: $M_{\rm spec, 1} \propto g_1\,R_1^2$ is obtained via the surface gravity $g_1$ and stellar radius $R_1$ derived from spectroscopy.  The stellar parameters and abundances are derived from a spectroscopic  analysis. We use the CMFGEN model atmosphere code \cite{Hillier1998}, though we perform the same analysis with two independent codes to explore systematics. The spectral analysis is provided in Supplementary Information.

All mass estimates agree within $1\sigma$. The final adopted value of $M_1$ is a weighted mean of $M_{\rm ev, 1}$ and $M_{\rm spec, 1}$ (we omit $M_{\rm SpT, 1}$, since it is not independent of  $M_{\rm ev, 1}$, being a mean of evolutionary masses).
 We note that the evolutionary masses are derived assuming single-star evolution models, while the properties of VFTS~243 suggests that the primary accreted mass from the black-hole progenitor (see below). The reaction of the star to the mass accretion is not trivial, but may result in a slightly different mass-luminosity relation compared with single stars. Another factor we neglect is the impact of rotation and microturbulence pressure on the derivation of the surface gravity, which should result in a slight underestimation of the true $\log g$ (by $\approx 0.1\,$dex). Instead of adopting a series of non-trivial assumptions, we consider a very generous error margin of $5\sigma$ (i.e., $M_1 = 25\pm12\,M_\odot$) when interpreting our results, and show that  a black-hole secondary cannot be avoided, regardless of the adopted primary mass (see also light curve discussion).

\subsection*{Spectral disentangling}

Faint companions in spectroscopic binaries may  be uncovered through various methods, such as carefully inspecting the spectra, subtracting a model for the primary, or cross-correlating the spectra with template models for the unseen secondary. Here, we use the method of spectral disentangling, which makes use of all available data (hence boosting the S/N to effectively $\approx 350$), and does not rely on pre-defined templates for the primary or secondary. Spectral disentangling is a method used to separate the component spectra of spectroscopic binaries while constraining the binary orbital solution. Few methods exist to disentangle spectroscopic binaries, operating either in Fourier space or wavelength space.
Our main method of choice is the shift-and-add technique \cite{Marchenko1998}. The shift-and-add technique generally yields comparable results to other techniques such as Fourier disentangling \cite{Shenar2020LB1}, and in some cases, the shift-and-add algorithm appears to be more robust \cite{Bodensteiner2020HR}. We also compare our results with those obtained from Fourier disentangling \cite{Hadrava1995}, and reach the same conclusions. 
This method was successfully used to identify the hidden companions in the black-hole imposters LB-1 and HR~6819 \cite{Shenar2020LB1, Bodensteiner2020HR} and in SB1 binaries (Mahy et al.\ 2022, A\&A, submitted; Shenar et al.\ prep.).

The original  shift-and-add algorithm assumes prior knowledge of the RVs of both components, $RV_{\rm 1, i}$ and $RV_{\rm 2, i}$, where $i \in {1, ..., N}$ runs over all available epochs of observation.  The computed spectra for the primary and the secondary of the $j^{\rm th}$ iteration, $A_{\rm j}$ and $B_{\rm j}$, are computed by shifting-and-subtracting the previous approximations $A_{\rm j-1}, B_{\rm j-1}$ from all observations, respectively, and then co-adding the observations using the available  RVs. The first iteration assumes a flat spectrum for the secondary. To accelerate the convergence, we  enforce the disentangled spectra to lie below the continuum in continuum regions \cite{Shenar2019, Quintero2020}. Convergence is reached within tens to hundreds of iterations, depending on the spectral line.

For VFTS~243, the RVs of the secondary are not constrained and the potential presence of a faint companion might bias the  primary RV measurements. 
In this case, one can perform the disentangling across the $K_1, K_2$ axes  (grid disentangling), where $K_1, K_2$ are the semi-amplitudes of the RV curves of both components. One may then disentangle specific spectral lines (or sets of lines), and  compute:

\begin{equation}
   \chi_{\rm reduced}^2(K_1, K_2) = \frac{1}{N_{\lambda} (N_{\rm epochs} - 2)}\,\sum_{i=1}^{N_{\rm epochs}}\,\sum_{k=1}^{N_{\lambda}} \frac{ \left(A_{i,k} + B_{i,k} - O_{i,k}\right)^2}{\sigma_i^2},
\label{eq:chi2}
\end{equation}
where $N_{\lambda}$ is the number of wavelength bins in the selected range, $N_{\rm epochs}$ is the number of epochs (37). $A_{i,k}, B_{i,k}$ are the disentangled spectra obtained for $K_1, K_2$, evaluated at $\lambda_k$, and  shifted to the appropriate RVs of the $i^{\rm th}$ epoch, $O_i$ is the observed $i^{\rm th}$ spectrum, and $\sigma_i$ is the S/N in the continuum region in the vicinity of the spectral line. The number of degrees of freedom is given by $ N_{\lambda} (N_{\rm epochs} - 2)$, since each pixel in each of the two disentangled spectra is considered a free variable.  The final disentangled spectra are then retrieved for the derived $K_1, K_2$ values. The spectra are retrieved up to a scaling factor which depends on the light ratio of the stars. The light ratio has no impact on the disentangling procedure. By minimising $\chi^2$, one can retrieve the $K_1, K_2$ values that best reproduce the data, and separate the composite spectra with these values.  As $K_2$ cannot be derived here, we adopt $K_2 = 3 \times K_1 = 244\,\kms$. The results are described in the main text and shown in Fig.\,\ref{fig:DisData}. The disentangled spectrum of the secondary is flat within the S/N (with the exception of the diffuse interstellar band at $\approx 4420\,$\AA~and cross-contamination from nebular lines).  This conclusion is independent of the choice of $K_2$ (see also Supplementary Information).

The spectra of VFTS~243 suffer from modest nebular contamination, which impacts primarily the strong Balmer lines. We account for the nebular contamination by extending the shift-and-add algorithm to three components, where the third component is static and is meant to represent the nebular lines \cite{Abdul-Masih2020Peanut}. Not accounting for nebular lines may result in spurious features, which  become significant in the amplified spectrum of the secondary. However, given the modest nebular contamination in VFTS~243, the impact of the nebular lines is negligible in all lines but H$\gamma$, and our conclusions remain unchanged regardless of the treatment of nebular lines. Moreover, nebular lines may result in false positives through spurious features in the disentangled spectrum, but not to a featureless spectrum, as observed for the secondary in VFTS~243.

\subsection*{Simulations of binaries hosting non-degenerate companions}

We perform extensive simulations to reject the presence of non-degenerate companions (Fig.\,\ref{fig:Sims}), exploring companion masses as low as our $5\sigma$ threshold ($6.5\,M_\odot$).  We consider binaries with main sequence companions, a binary with a helium-star companion, and a triple scenario in which the secondary is a binary in itself. Main sequence companions are modelled with the TLUSTY \cite{Hubeny1995, Lanz2007} model atmosphere grids, while helium stars are modelled using the Potsdam Wolf-Rayet (PoWR) model atmosphere code \cite{Hamann2003, Sander2015}.  As a final test, one of our team members (KEB) created "blind simulations", which were disentangled without prior knowledge of their input parameters. In all of the above cases, the companion could be retrieved without ambiguity.  A detailed account of the simulations is provided in Supplementary Information. 

\section*{Photometric variability}

The light curve of VFTS~243 contains evidence of both short- and long-timescale variability, though the long-term variability does not appear to be of stellar origin and is removed from the data.  A Fourier analysis of the $I$-band light curve shows a peak at a period of 5.201 days, which  coincides exactly with half the spectroscopic orbital period. The phased light curve shows a modulation consistent with the presence of ellipsoidal variations, with a semi-amplitude of $A_{\rm ellipsoidal} = 0.0015 \pm 0.0003$.  The ellipsoidal variability amplitude at fixed period depends primarily on the density of the tidally distorted O star (hence $M_1, R_1$) and on the inclination $i$, with weaker dependencies on the mass ratio ($M_2/M_1$) and atmospheric parameters of the star.  To explore the range of binary parameters that can match the observed amplitude, we calculate a set of model light curves of binaries satisfying the orbital constraints that comprise a dark companion and an O star radius of 10.5\,$R_{\odot}$ using the PHOEBE tool \cite{Prsa2005}.   

All plausible values of $M_1$ imply a companion mass $M_2\gtrsim 9\,M_{\odot}$.  The lower the value  of $M_1$, the more tidally distorted the primary becomes. In this case, the observed low-amplitude variability can only be matched for increasingly lower inclinations, which in turn imply higher secondary masses. It is therefore interesting to note that while a lower primary mass reduces the minimum  mass of the secondary, it  increases the actual mass of the secondary required to reproduce the ellipsoidal variability amplitude. For example, if the mass of the primary were only $12\,M_\odot$ (instead of the derived $25\,M_\odot$), the light curve would require an orbital inclination smaller than $30^\circ$ and the secondary would weigh $\approx 20\,M_\odot$, making it impossible to hide in the data unless it were a black hole. The light curve shows that the black-hole interpretation is inescapable, even in the case of an extreme overestimation of the primary mass below the 5$\sigma$ threshold explored here.

\section*{Radial velocity and proper motion}

Next to the vanishing eccentricity, another indication for lack of a supernova kick can be obtained from the relative motion of VFTS~243 with respect to its natal environment. The systemic velocity of VFTS~243, $\Gamma = 260.2\pm 0.9\,\kms$, is within 1$\sigma$ of the measured mean for the Tarantula, $\langle \Gamma \rangle = 271.6\pm12.2\,\kms$ \cite{Evans2015}. We note that the standard deviation of $12.2\,\kms$ likely represents a true dispersion rather than a measurement error (which is typically a factor 10-20 smaller). Estimates on the proper motion can be obtained from {\it Gaia} \cite{Gaia2021}. We computed the means of the two components of the proper motion vector (RA, DEC) of all stars within 4' of VFTS~243 whose parallaxes and proper motions are consistent with the LMC.
Since the errors in this case are dominated by statistical errors, we simply assume that the RV dispersion reflects the dispersion in RA and DEC directions as well. We find $\langle \varv_{\rm RA} \rangle = 396 \pm 12\,\kms$ and $ \langle \varv_{\rm DEC} \rangle = 145 \pm 12\,\kms$ for the environment mean. The measured {\it Gaia} velocity components of VFTS~243 are $\varv_{\rm RA} = 408 \pm 8\,\kms$ and $\varv_{\rm DEC} = 143 \pm 12\,\kms$. Hence,  each component of the 3D velocity vector of VFTS~243 is within $1\sigma$ of the mean.  This gives further support for the lack of a strong kick associated with the core-collapse of the black-hole progenitor.

\section*{Data availability} 

The VFTS and TMBM spectra are available on the ESO archives (archive.eso.org/cms.html). The input files of our stellar evolution model (Supplementary Information) are available on \url{doi.org/10.5281/zenodo.6514645}. The OGLE light curve is available for download on \url{https://cdsarc.u-strasbg.fr/ftp/vizier.submit/OGLE_VFTS243}.
Any other processed materials  are available
from the corresponding author upon reasonable request.

\section*{Code availability}

The fd3 tool used for Fourier disentangling is available online (\url{sail.zpf.fer.hr/fdbinary/}). The shift-and-add Python implementation used here is available from the corresponding author upon reasonable request.
We refer the reader to the CMFGEN (\url{kookaburra.phyast.pitt.edu/hillier/web/CMFGEN.htm}), PoWR. (\url{www.astro.physik.uni-potsdam.de/PoWR}), FASTWIND (\url{fys.kuleuven.be/ster/research-projects/equation-folder/codes-folder/fastwind}), and PHOEBE (\url{http://phoebe-project.org/}) webpages for access and availability policies.

\section*{Acknowledgments}
This research has received funding from the European Research Council (ERC) under the European Union’s Horizon 2020 research and innovation programme (TS and HS, grant agreement number 772225: MULTIPLES). TS acknowledges support from the European Union's Horizon 2020 under the Marie Skłodowska-Curie grant agreement No 101024605.  This work is based on observations collected at the European Southern Observatory under program IDs 182.D-0222, 090.D-0323 and 092.D-0136.  LM thanks the European Space Agency (ESA) and the Belgian Federal Science Policy Office (BELSPO) for their support in the framework of the PRODEX programme. PM acknowledges support from the FWO junior postdoctoral fellowship No. 12ZY520N. We thank John Hillier for making the CMFGEN code available. CH acknowledges support from the KU Leuven Research Council (grant C16/17/007: MAESTRO). PAC acknowledges support from the UK Science and Technology Facilities Council research grant ST/V000853/1.  VHB acknowledges the support of the Natural Sciences and Engineering Research Council of Canada
(NSERC) through grant RGPIN-2020-05990. MG acknowledges support from the Ministry of Science and Innovation through a Europa Excelencia grant (EUR2020-112157).
The PoWR code, as well as the associated post-processing and
visualisation tool WRplot, has been developed under the guidance of
Wolf-Rainer Hamann with substantial contributions from Lars Koesterke,
G\"otz Gr\"afener, Andreas Sander, Tomer Shenar and other co-workers and
students. This work has received funding from the European Research Council (ERC) under the European Union’s Horizon 2020 research and innovation programme (Grant agreement No.\ 945806). This work is supported by the Deutsche Forschungsgemeinschaft (DFG, German Research Foundation) under Germany’s Excellence Strategy EXC 2181/1-390900948 (the Heidelberg STRUCTURES Excellence Cluster). ST acknowledges support from the Netherlands Research Council NWO (VENI 639.041.645, VIDI 203.061 grants). AH and DJL acknowledge support by the Spanish MCI through grant PGC-2018-0913741-B-C22 and the Severo Ochoa Program through CEX2019-000920-S. JMA acknowledges support from the Spanish Government Ministerio de Ciencia, Innovaci\'on y Universidades through grant PGC2018-095\,049-B-C22. We thank Isca Mayo and Sara Pinilla for their work on the visualisation of VFTS~243. We acknowledge the Cambridge Astronomical Survey Unit for providing access to the background image in Figure\,1. This research has made use of the SIMBAD database, operated at CDS, Strasbourg, France.

\section*{Author contributions}
TS, HS, LM and MF developed the analysis methodology. TS identified the target, performed the disentangling and simulations, derived the radial velocities, analysed the spectrum with PoWR, and wrote the manuscript. HS led the TMBM observing campaign, prepared the observations and, together with LAA, performed data reduction, variability and orbital analysis of the TMBM sample, and contributed to the interpretation. LM analysed the spectrum with CMFGEN. KEB performed the light-curve analysis, investigated the spectra independently, and produced mock simulations for blind-testing. PM and NL contributed to the  interpretation of the evolutionary status and X-ray properties of the system.   CH analysed the spectrum with FASTWIND. MF performed an independent test with Fourier disentangling.  KS contributed to the discussion of X-ray production. DJL contributed to the assessment of the light ratios of hypothetical main sequence stars and investigated the spectra independently. JMA provided an independent measure of the stellar parameters and luminosity. PM, AP, FS, and MG contributed to the computation of synchronisation and circularisation timescales. 
All other authors contributed to acquisition of the data, the discussion of the results and commented on the manuscript. 

\section*{Correspondence}
Correspondence and requests for materials should be addressed to Dr.\ Tomer Shenar (email: T.Shenar@uva.nl).

\section*{Competing Interests}
The authors declare no competing interests. 

\newpage


\newpage



\pagebreak

\section*{Supplementary Information}

\newcommand{\beginsupplement}{
\setcounter{figure}{0}
\renewcommand{\figurename}{Supplementary Figure}
\setcounter{table}{0}
\renewcommand{\tablename}{Supplementary Table}}
\beginsupplement

\subsection*{Data}

{\bf Spectroscopy:} The 32 epochs of spectroscopy that are the backbone of our study were obtained in the framework of the TMBM campaign. The broader TMBM campaign, the data acquisition and their reduction are described in \cite{Almeida2017}.  Here, we briefly describe the main elements relevant for the present paper. The optical light from VFTS~243, along with that of about 100 other scientific targets in the field of view, was picked up by MEDUSA fibers appropriately positioned in the focal plate of the instrument and fed into the GIRAFFE spectrograph. In addition, 14 sky fibers were positioned on blank regions across the field-of-view and were used to obtain a median sky and nebular spectrum at each epoch. The MEDUSA fibers have a 1.2" entrance diameter. The GIRAFFE data were acquired  with the L427.2 (LR02) grating, which covers the wavelength range 3964--4567\,\AA~at a resolving power of $R = 6400$. 

We also use five additional LR02 spectra from the VFTS campaign, described in \cite{Evans2011}, amounting to a total of 37 spectra. Each TMBM  (resp.\ VFTS) observation consisted of three 900\,s exposures (resp.\ two 1815\,s exposures) taken back-to-back. The data were reduced in a similar way using the ESO Common Pipeline Library (CPL)  and calibrations obtained in the mornings following the night-time observations of each epochs.  The raw data were bias-subtracted, flat-fielded and wavelength calibrated using a standard approach \cite{Evans2011, Almeida2017}. The individual spectra corresponding to science and sky fibers were extracted in SUM mode. The science spectra were then sky subtracted using the median sky spectrum. Finally, the spectra were normalized following a semi-automatic iterative procedure described in \cite{Sana2013}. The GIRAFFE wavelength calibration precision achieved thanks to the reference ThAr calibration frames is of $\sim300$~m\,s$^{-1}$, and is consistent from one epoch to another. 

The reduced data have a mean S/N of 61 per pixel.  
For the spectral analysis, we use the disentangled spectrum of the primary (which is very similar to the co-added spectrum) in the range 3964--4567\,$\AA$, amounting to a total S/N of $\approx 350$. We extend this spectrum with VFTS spectra acquired with the LR03 and HR15N gratings of FLAMES, covering 4499–5071$\,\AA$  and 6442–6817$\,\AA$.

\noindent
{\bf Photometry:} A light curve of VFTS~243 was obtained from  the Optical Gravitational Lensing Experiment (OGLE) data reductions III and IV.   We refer the reader to \cite{Udalski2003, Udalski2008, Udalski2015} for detailed information on the data acquisition, calibration, and reduction.  For the spectral energy distribution, we used photometry in the F275W, F336W, F555W,  F658N, F775W, F110W, and F160W filters of the Hubble Space Telescope's (HST) Wide Field Camera 3, UVIS and IR channels \cite{Sabbi2016},  and far UV B1, B5 fluxes measured with the Ultraviolet Imaging Telescope (UIT) \cite{Parker1998}. A compilation is provided in Supplementary Table\,\ref{tab:Photometry}.

\begin{table}[!h]
\centering
\caption{Photometry used in our study. Listed are the bands and their effective wavelength, measured magnitudes and corresponding photometric system (Vega or AB), epochs of observation, and references. The measurements have varying errors of the order of  0.01-0.02\,mag, but we adopt a constant error of 0.02\,mag due to possible photometric variability (see Sect.\ Photometric variability).  }
\resizebox{.7\textwidth}{!}{\begin{tabular}{cccc}\hline \hline
\vspace{-4mm}\\ 
 Band &   Measurement (mag) & Epoch & Reference   \\
\hline
UIT B1 ($1521\,\AA$) &  14.04 (AB)	& 1990-12-05 &	\cite{Parker1998}\\
UIT B5 ($1615\,\AA$) &  13.96 (AB)	& 1990-12-05 &	\cite{Parker1998}\\
F275W ($2720\,\AA$)  & 14.44 (Vega) & 2013-03-28 & \cite{Sabbi2016} \\
F336W ($3360\,\AA$) & 14.36 (Vega) & 2013-03-28 & \cite{Sabbi2016} \\
F555W ($5235\,\AA$)  & 15.33 (Vega) & 2013-09-26 & \cite{Sabbi2016} \\
F658N ($6586\,\AA$) & 15.00 (Vega) & 2012-12-17 & \cite{Sabbi2016} \\
F775W ($7610\,\AA$) & 14.97 (Vega) & 2011-10-29 & \cite{Sabbi2016} \\
F110W ($1.083\,{\mu}m$)  & 14.85 (Vega) & 2013-05-22 & \cite{Sabbi2016} \\
F160W ($1.528\,{\mu}m$) & 14.65 (Vega) & 2013-06-03 & \cite{Sabbi2016} \\
\hline
\end{tabular}}
\label{tab:Photometry}\end{table}

\subsection*{Revised orbital analysis}

While the orbit of VFTS~243 has been previously constrained \cite{Almeida2017}, we derive it independently here following a different radial-velocity (RV) measurement technique that is less sensitive to the presence of a putative secondary signature. We measure the RVs of the He\,{\sc i}\,$\lambda\lambda$~4026, 4388, 4471 and He\,{\sc ii}\,$\lambda\lambda$~4200, 4541 lines  by fitting a parabola to the maximum of the cross-correlation functions (CCF) computed using a template \cite{Zucker1994}. Initially, we use a suitable synthetic spectrum from our spectral modelling (see below). We then measure the RVs, and co-add the spectra using these measurements to create a wavelength-calibrated, co-added, high S/N spectrum as our final template \cite{Shenar2019}. This template is then used to re-measure the RVs. To avoid an underestimation of the error due to auto-correlation of the noise in the co-added template \cite{Marsh1994}, we fit the parabola to the maximum of the CCF over a range that is significantly wider than the few $\kms$~range in which auto-correlation could play a role. We note that the errors obtained when using a synthetic spectrum are comparable to those obtained when using the co-added template (difference of means smaller than 1\%).

 We implement a Markov-chain Monte Carlo (MCMC) simulation using Python's  emcee package \cite{Foreman-Mackey2013} to derive an orbital solution and obtain robust errors on the orbital parameters by fitting

\begin{equation}
\label{eq:RVorbit}
    {\rm RV}(\nu) = \Gamma + K_1\,\left( \cos(\omega + \nu) + e \cos \omega \right),
\end{equation}
where $\nu$ is the true anomaly. We perform the simulations for RVs measured using only He\,{\sc i}, only He\,{\sc ii}, and all lines. The three sets of results agree within their respective errors, and so we adopt the results obtained when using all He lines (provided in Table\,1 in the main text)  and  shown in Supplementary Figs.\,\ref{fig:orbit} and \ref{fig:Corner}. The derived orbital elements agree well with those previously derived \cite{Almeida2017}, but provide robust error estimates, which are critical especially for the eccentricity: $e = 0.017^{+0.014}_{-0.011}$ (68\% confidence interval). For comparison, we obtain $e = 0.011^{+0.012}_{-0.008}$ using only He\,{\sc i} RVs, and $e = 0.027^{+0.017}_{-0.015}$ using only He\,{\sc ii} RVs.

\begin{figure}
\centering
\includegraphics[width=\textwidth]{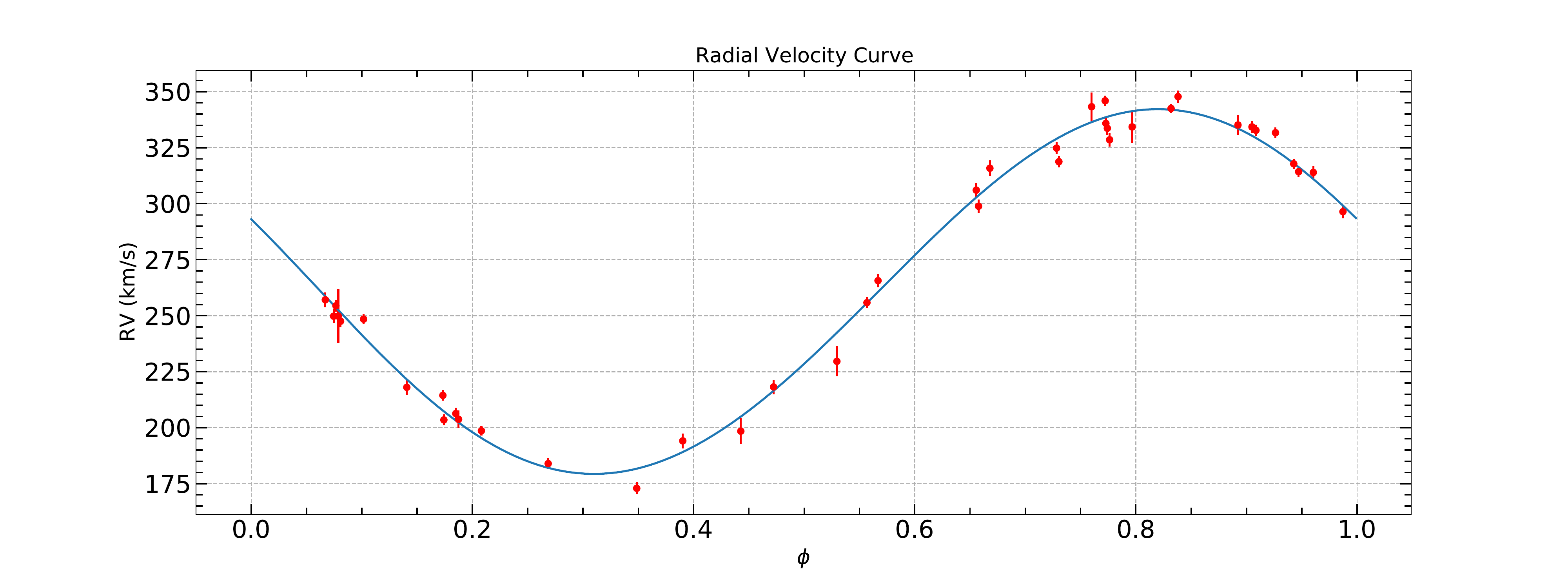}
\caption{{\bf RVs of the O7~V primary star (symbols) measured from He\,{\sc i, ii} lines.} Errors represent 1$\sigma$ measurement errors. The best-fit RV-curve has been overlaid (plain line).} \label{fig:orbit}
\end{figure}

\begin{figure}
\centering
\includegraphics[width=\textwidth]{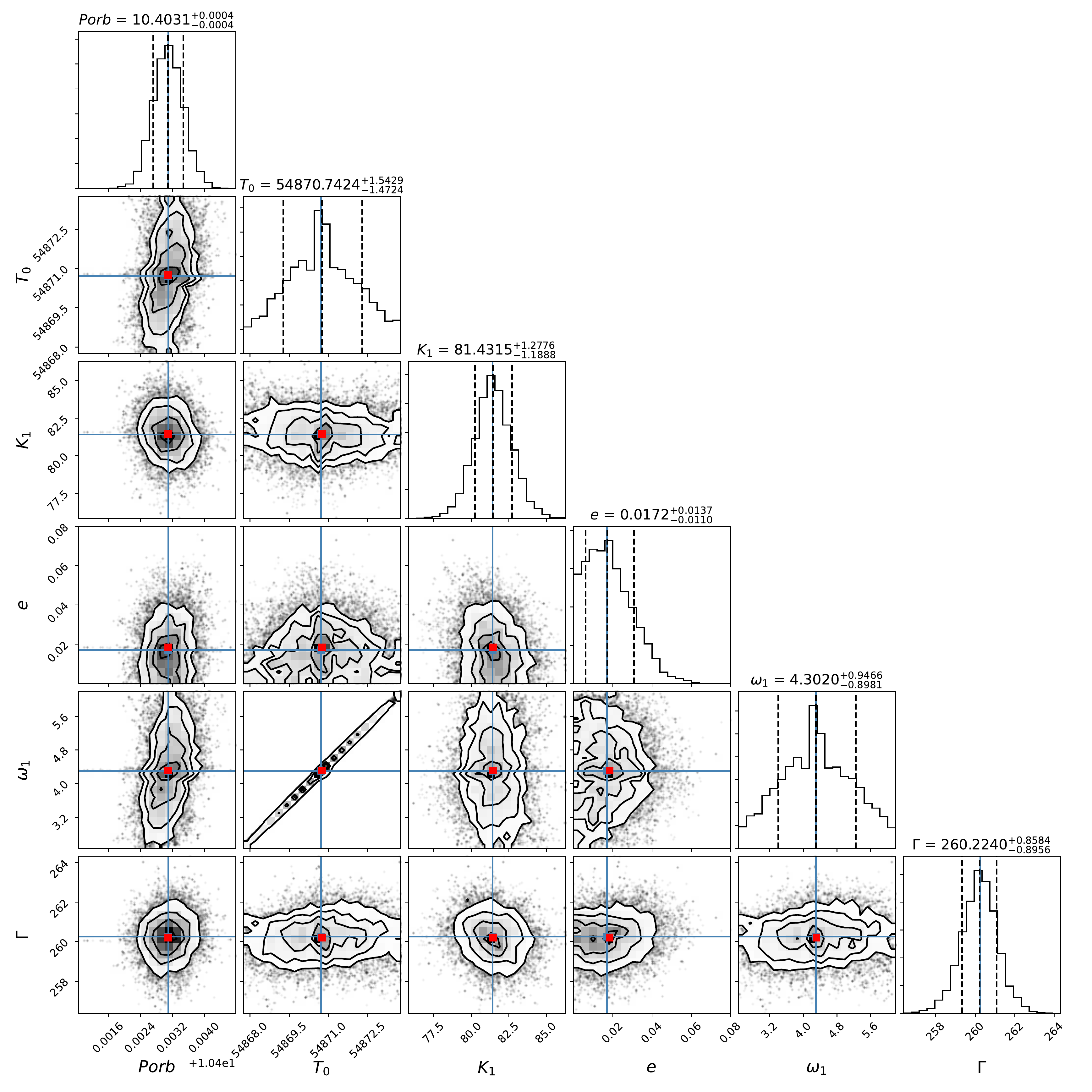}
\caption{{\bf Corner plot for the five orbital parameters.} Results are obtained from our Monte Carlo simulations using RVs measured from He\,{\sc i} and {\sc ii} lines combined. Errors represent the 68\% confidence intervals. We note that $\omega = \omega_1 - \pi$ is provided (in degrees) in Table\,1 in the main text, in agreement with Eq.\,(\ref{eq:RVorbit}) and the definition in the TMBM campaign \cite{Almeida2017}. } \label{fig:Corner}
\end{figure}

\subsection*{Spectral analysis}

The projected rotational velocity $\varv \sin i$ was derived using the Fourier method introduced by \cite{Gray1992}, implemented by the  IACOB-BROAD tool \cite{Simon-Diaz2007}. The mean  obtained from He\,{\sc i}\,$\lambda 4026, 4388$, $4471$, $4713$, and $4922$ is $\varv \sin i = 181\pm16\,\kms$, where the error is taken from the Nyquist frequency of the  spectral resolving power, and the measurements have a standard deviation of $6.7\,\kms$. The tool also provides various measures for   the macroturbulent velocity $\varv_{\rm mac}$. However, given that the results are highly line-dependent ($30 - 190\,\kms$), we simply fix $\varv_{\rm mac} = 60\,\kms$ judging by the fit quality. 

The remaining atmospheric properties reported in our study were derived primarily using the CMFGEN code \cite{Hillier1998}, which solves the radiative transfer in expanding atmospheres while relaxing the assumption of local thermodynamic equilibrium (non-LTE). However, we also performed independent analyses with the PoWR and FASTWIND codes (see below). 
We assume a microturbulent velocity of $\varv_{\rm mic} = 10\,\kms$ and a standard $\beta$-law with $\beta=1$ for the wind velocity field.
To derive the effective temperature $T_{\rm eff}$ and surface gravity $\log g$ of VFTS~243, we performed a $\chi^2$ analysis based on a grid of synthetic spectra computed with CMFGEN covering from 27.0~kK to 47.0~kK in $T_{\rm eff}$ and from 3.1 to 4.4 in $\log g$, with steps of 1.0~kK and 0.1 dex, respectively. The abundances (including hydrogen) and luminosity were set to the predictions used for the Brott models \cite{Brott2011}.
The terminal wind speed was fixed to $\varv_\infty = 2100\,\kms$ (\cite{Brands2022} and Hawcroft et al., in prep.). Lacking constraints on clumping from the UV, we assume a smooth wind (clumping factor $D=1$), and provide the unclumped mass-loss rate $\dot{M} \sqrt{D}$.

Each model of the grid was first convolved with rotational profiles, mimicking the projected rotational velocity, then convolved with a radial-tangential profile mimicking the macroturbulence. Finally,  to account for the finite instrumental resolution, a third convolution  with Gaussian profiles with a full-width half maximum of $\Delta \lambda = 0.7\,$\AA~(corresponding to $R\approx 6\,000$) was performed.  The effective temperature was constrained from the ionisation balance between the He~{\sc i} and  He~{\sc ii} lines. We used the  He~{\sc i+ii}~4026,  He~{\sc i}~4143,  He~{\sc i}~4388,  He~{\sc i}~4471,  He~{\sc ii}~4200 and He~{\sc ii}~4542 lines. The surface gravity was obtained from the wings of the H$\delta$, H$\gamma$, H$\beta$, and H$\alpha$ lines. We took care of removing the cores of the H$\alpha$ and H$\beta$ lines as they are contaminated by nebular emission. We note that  the nebular emission barely impacts the helium lines. The best estimates for VFTS~243 are $\teff{} = 36.0 \pm 1.0$~kK and $\log~g = 3.7 \pm 0.1$~[cgs]. The fit to the data is shown in Supplementary Fig.\,\ref{fig:specan}.

Once the effective temperature and surface gravity derived, we took the corresponding model to fit the spectral energy distribution of VFTS~243. We use the extinction law described by \cite{Maiz2014} for stars in 30~Dor region, and the distance modulus of the LMC (DM = 18.495 \cite{Pietrzynski2013}).  We assume a constant error of 0.02\,mag on all photometry measurements, which is  comparable to the measurement errors (typically 0.01-0.02\,mag) and possible photometric variability (see Sect.\ Photometric variability).  We obtain a stellar luminosity of $\log(L/L_{\odot}) = 5.20 \pm 0.04$. Finally, we re-compute a set of CMFGEN models by adopting the effective temperature, surface gravity, and stellar luminosity and only allowing the nitrogen content to vary. Focusing on the N~{\sc iii} quadruplet at $\sim 4515$~\AA, we derived a nitrogen number fraction of $\epsilon_{\rm N} = 8.1$  (where $\epsilon_i = 12 - \log(n_{\rm H}/n_i)$, and $n_i/n_{\rm H}$ is the number ratio of the respective element relative to hydrogen.) , amounting to a mass fraction of $X_{\rm N} = 0.11 \pm 0.05\%$ (16 times the baseline value). This indicates that the primary component is strongly enriched in nitrogen at its surface, implying the presence of CNO-processed material in its atmosphere. No clear indications for carbon depletion are seen, but  a factor $\approx 2$ depletion is consistent within errors.

\begin{figure}
\centering
\includegraphics[width=\textwidth]{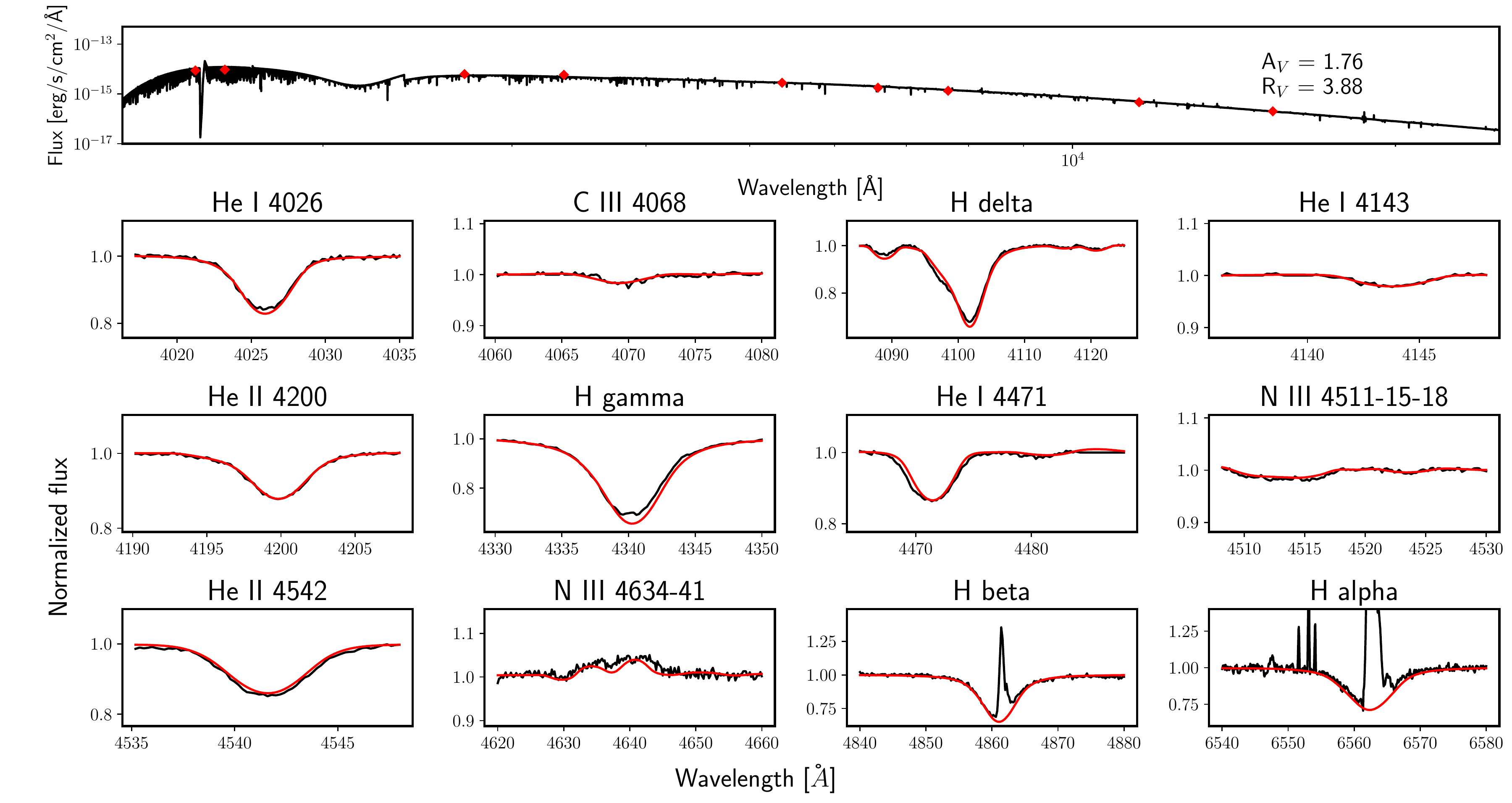}
\caption{{\bf Best-fitting CMFGEN atmosphere model.} The spectral energy distribution (upper panel) and selected spectral line profiles (lower panel) are shown. The relative and total extinctions in the $V$-band $R_V$ and $A_V$ are noted. } \label{fig:specan}
\end{figure}

\subsection*{Simulations of binaries hosting non-degenerate companions}

Evidently, the disentangled spectrum of the secondary contains no features that can be considered to be of stellar origin. In Supplementary Fig.\,\ref{fig:MultiK2}, we illustrate that this conclusion is obtained regardless of the secondary's RV amplitude $K_2$ chosen for disentangling. To show that the companion cannot be a faint non-degenerate source,  we performed extensive simulations to reject the presence of non-degenerate companions, exploring companion masses as low as our $5\sigma$ threshold ($6.5\,M_\odot$).  We describe the simulations below. For simplicity, and to mimic a realistic situation in which $K_2$ is uncertain, we always disentangle the spectra for a fixed value of $K_2 = 300\,\kms$, regardless of the input parameters.

\begin{figure}
\centering
\includegraphics[width=\textwidth]{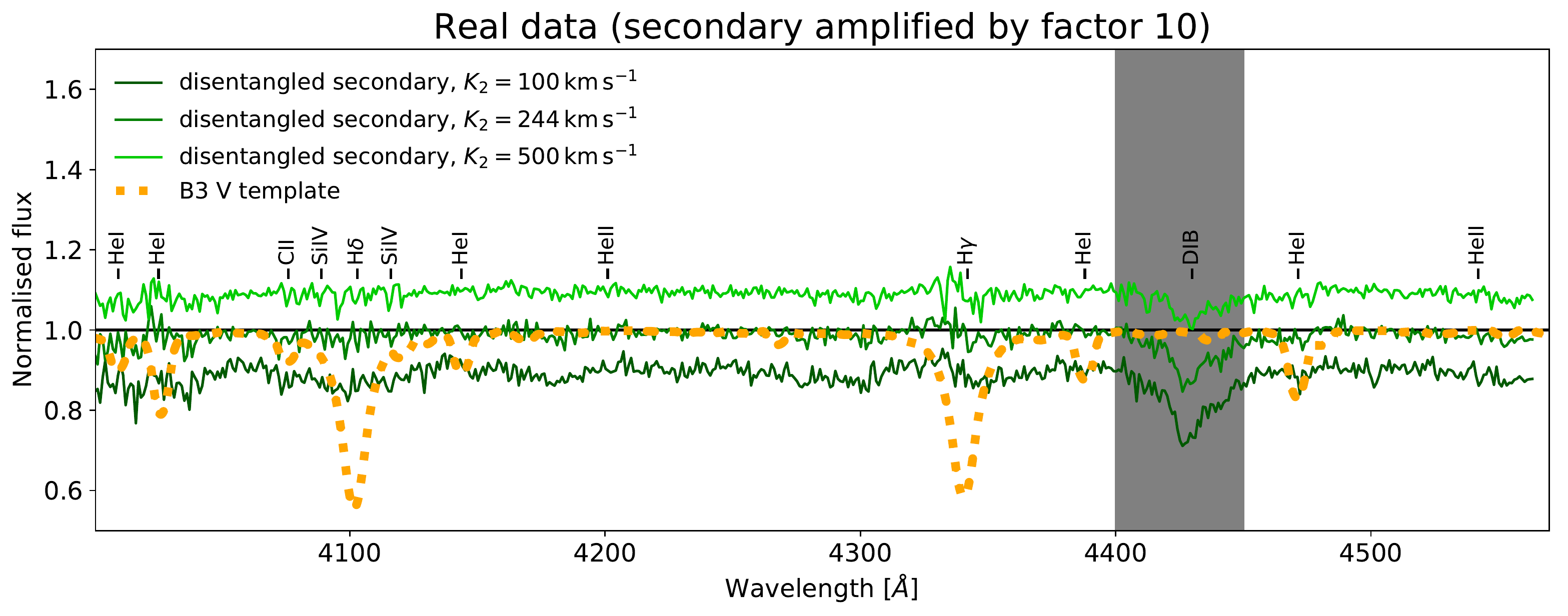}
\caption{{\bf Comparison between a B3~V template and the disentangled spectra of the secondary.} The comparison is shown  for three distinct $K_2$ values: $K_2 = 500$ (top), 244 (middle), and $100\,\kms$, illustrating the very weak dependence of the disentangled spectra on $K_2$. The spectra are binned at $\Delta \lambda = 1\,\AA$ for clarity.   } 
\label{fig:MultiK2}
\end{figure}

\noindent
{\bf O7~V + B1~V:} Our minimum mass is $M_{\rm min, 2} = 8.7\,M_\odot$, which corresponds roughly to a B1~V dwarf \cite{Harmanec1988} moving with  $K_2 = K_1\,M_1/M_2 = 230\,\kms$ for a main sequence companion. Using the BONNSAI tool with this mass and the estimated single-star  age of 3.9\,Myr \cite{Schneider2018} as input, this corresponds roughly to a secondary with $T_{\rm eff, 2} = 25\,$kK and $R_2 = 3.6\,R_\odot$. We estimate the $B$-band  magnitude of the components using corresponding models available from the non-LTE plane-parallel model atmosphere code TLUSTY
\cite{Hubeny1995, Lanz2007}, 
and find  $M_{B, 2} = -2.1\,$mag. With $M_{B, 1} = -5.1\,$mag from our model atmosphere of the primary, this amounts to a $B$-band flux contribution of $l_2 = 5.9\%$ for the secondary.
For the spectrum of the primary, we use the disentangled spectrum as template. For the secondary, we use a corresponding TLUSTY spectrum.  Finally, we assume a conservatively large projected rotational velocity of $\varv_2 \sin i = 300\,\kms$~for the secondary and convolve its spectrum accordingly, making it harder to detect in the spectrum. We create 37 mock spectra with the individual S/N values and phasing of the real data. The companion is easily retrieved.

\noindent
{\bf O7~V + B3~V:} We follow the same steps as for the B1~V simulations, but this time for the $5\sigma$ lower threshold of $M_2 \approx 6.5\,M_\odot$. This corresponds roughly to a B3~V star with $K_2 = 335\,\kms$. Such a star is expected to have $T_{\rm eff, 2} = 21\,$kK and $R_2 = 2.7\,R_\odot$, corresponding to  $M_{B, 2} = -1.3\,$mag and a flux contribution of merely $l_2 = 2.9\%$. We use a corresponding TLUSTY model with  $\varv_2 \sin i = 300\,\kms$ for the secondary. Despite its low flux contribution, the companion is readily retrieved.

\begin{figure}
\centering
\includegraphics[width=\textwidth]{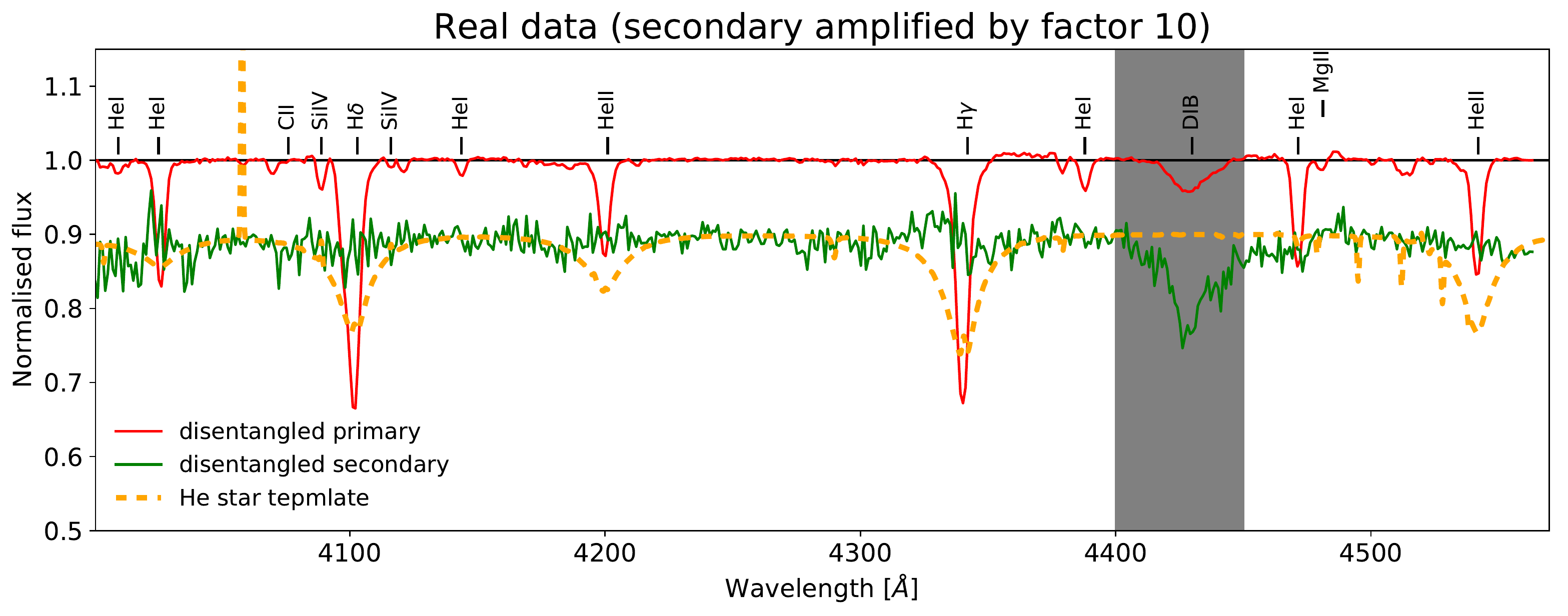}
\caption{{\bf Comparison of the disentangled spectra with a helium-star template.} As Fig.\,4 in the main text, but comparing the disentangled spectrum of the secondary to a helium-star template computed with the PoWR code. Evidently, the presence of such a star can be rejected from the data. } 
\label{fig:disenHe}
\end{figure}

\noindent
{\bf O7~V + He:} Instead of a main sequence star,  the companion might be a stripped helium star \cite{Gies1998, Wang2018}, although the likelihood for such system is roughly ten times lower than for a black-hole companion \cite{Shao2021}. Here, we consider a He star with a mass as low as $M_2 = 6.5\,M_\odot$ to explore the most extreme scenario corresponding to a 5$\sigma$ lower limit on the companion mass. We adopt parameters from \cite{Goetberg2018} ($T_{\rm eff} = 80\,$kK, $R_* = 1.1\,R_\odot$) with $\dot{M}$ from \cite{Vink2017} and use the PoWR code \cite{Hamann2003, Sander2015}  to compute a corresponding synthetic spectrum. The star is expected to contribute a mere $l_2 \approx 3\%$ to the flux in the $B$ band. Nevertheless, we can unambiguously uncover its signature in the disentangled spectra Supplementary Fig.\,\ref{fig:disenHe} illustrates further that a helium star cannot be present in the data.
 
\noindent
{\bf O7~V + (B9~V+B9~V):} Another possibility is that the companion is in itself a binary \cite{Mazeh2020}, and the system is hence a triple. Since the luminosity sharply drops with mass, the most challenging inner-binary configuration is one in which both components are of equal mass. Fitting a binary inside a 10.4\,d  orbit requires substantial fine-tuning. If the system is to remain stable, the inner period cannot be larger than $\approx 2\,$d. For such a configuration,  the Kozai-Lidov timescale is of the order of $100\,$d, and further tuning of  the mutual inclination would be needed to avoid a merger of the inner two stars \cite{Toonen2020}. Nevertheless, we consider two stars with $M_{\rm 2,A} = M_{\rm 2, B} = 3.2\,M_\odot$, which corresponds to our lowest $5\sigma$ threshold. We consider main sequence stars, since any other scenario can be ruled out from an evolutionary perspective. This configuration corresponds roughly to O7~V + (B9~V+B9~V). We use approximate TLUSTY models for the companions, this time with $T_{\rm eff} = 15\,$kK and $\log g = 4.0\,$[cgs] (which are at the edge of the grid).

Mass-luminosity calibrations suitable for main-sequence dwarfs  \cite{Harmanec1988} imply that each B9~V component contributes a mere $0.7\%$ to the total flux. We assume that the stars are on a 1.3\,d circular orbit, have $K_{\rm 2,A} = K_{\rm 2, B} = 150\,\kms$ each (which implies $K_2 = 342\,\kms$ for the center of mass of the secondary), and assume a projected rotation of $300\,\kms$ for each. We disentangle the mock data as if this were a binary with $K_2 = 300\,\kms$.  While faint, the Balmer lines immediately reveal the presence of non-degenerate companions. The data are thus not even consistent with a binary abiding to our 5$\sigma$ lowest $M_2$ threshold.

\noindent
{\bf ``Blind" testing with non-idealised mock data:} As a final test, one of our team members (KEB) created mock data of four binaries by injecting various template spectra into the original data, while adding additional random noise to each spectrum. A summary of the simulations is provided in Supplementary Table\,\ref{tab:Sims}.  The mock data were disentangled  without prior knowledge of their  properties, with the goal of identifying the OB+BH binary (simulation D). The results are shown in Supplementary Fig.\,\ref{fig:SimsBlind}. The presence of a companion in all simulations but simulation D is readily evident. 

\begin{table}
\centering
\caption{Summary of the input parameters of the secondaries (effective temperatures $T_{\rm eff}$, surface gravities $\log g$, projected rotational velocities $\varv \sin i$, and light contributions $l_2$) used to produce the four ``blind simulations" by an independent team member (KEB).}
\resizebox{.45\textwidth}{!}{\begin{tabular}{lcccc}\hline \hline
Simulation & A & B & C & D   \\
\hline 
$T_{\rm eff}\,$[kK] & 22 & 22 & 25 & -   \\
$\log g\,$[cgs] & 4.0 & 4.0 & 4.0 & -   \\
$\varv \sin i\,[\kms]$ & 300 & 50 & 300 & -   \\
$l_2$ & 0.03 & 0.03 & 0.06 & 0   \\
\hline
\end{tabular}}
\label{tab:Sims}\end{table}

\begin{figure}[!h]
\centering
\includegraphics[width=\textwidth]{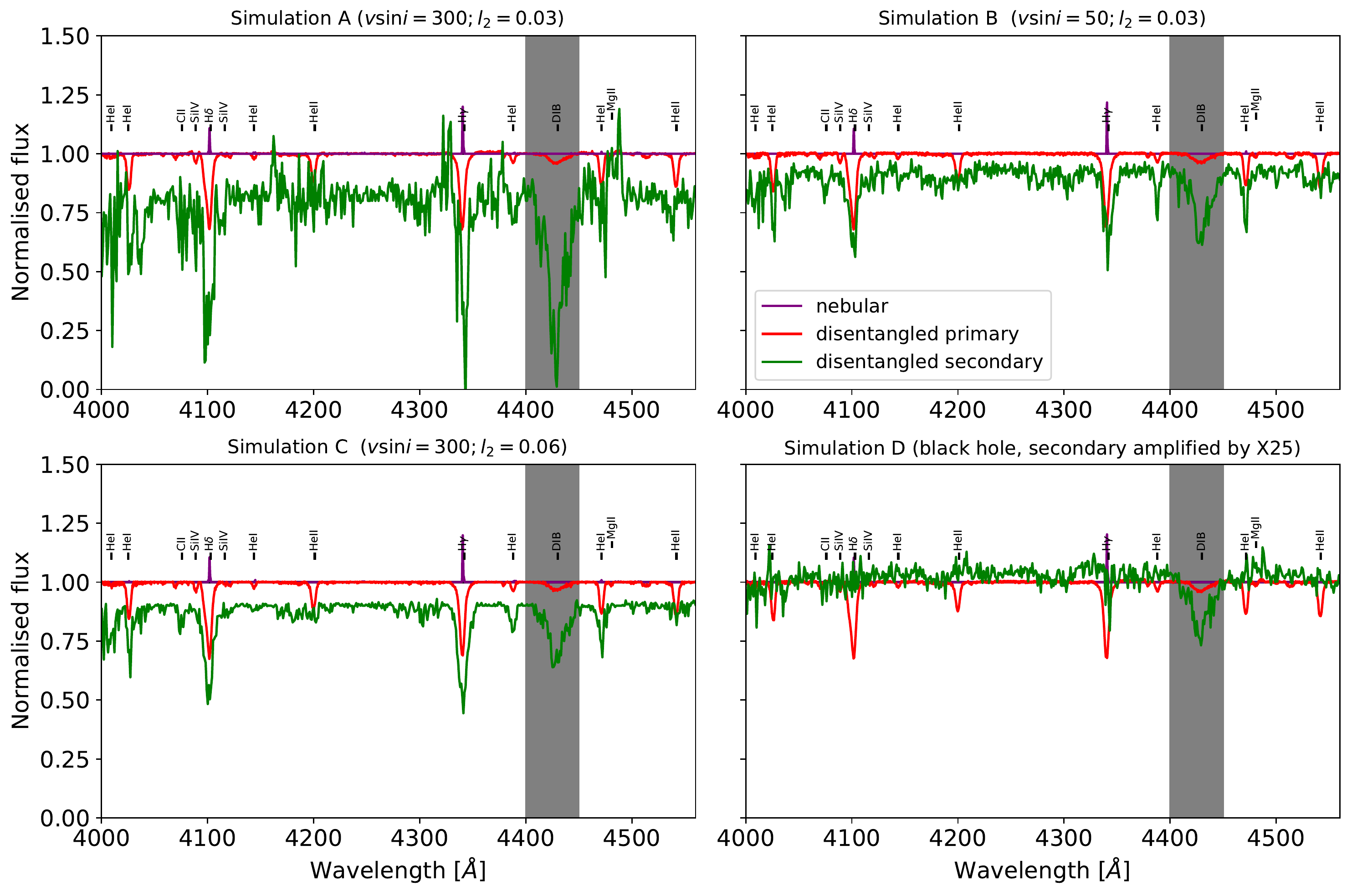}
\caption{{\bf Disentangled spectra of four ``blind" simulations.} The simulations include the injection of synthetic spectra and additional noise in the real data, with companion flux contributions as low at 3\% (see Supplementary Table\,\ref{tab:Sims}). The OB+BH simulation (D) is readily evident.  } \label{fig:SimsBlind}
\end{figure}


\section*{Photometric variability}

VFTS 243 is regularly observed by the OGLE project \cite{Udalski2003, Udalski2008, Udalski2015}. The source has two entries in the OGLE-IV database, LMC517.01.10094 and LMC553.25.24167, because it is located in the overlapping parts of two adjacent fields. We analyze the light curve of LMC553.25.24167, which is the less noisy of the two, probably because LMC517.01.10094 is located close to the edge of its field. The $I-$band light curve contains 1073 photometric points obtained over a 19 year period from 2001 to 2020, with a median magnitude $I=14.92\,$mag and typical photometric uncertainty of 0.005\,mag.

\begin{figure}[!h]
\centering
\includegraphics[width=\textwidth]{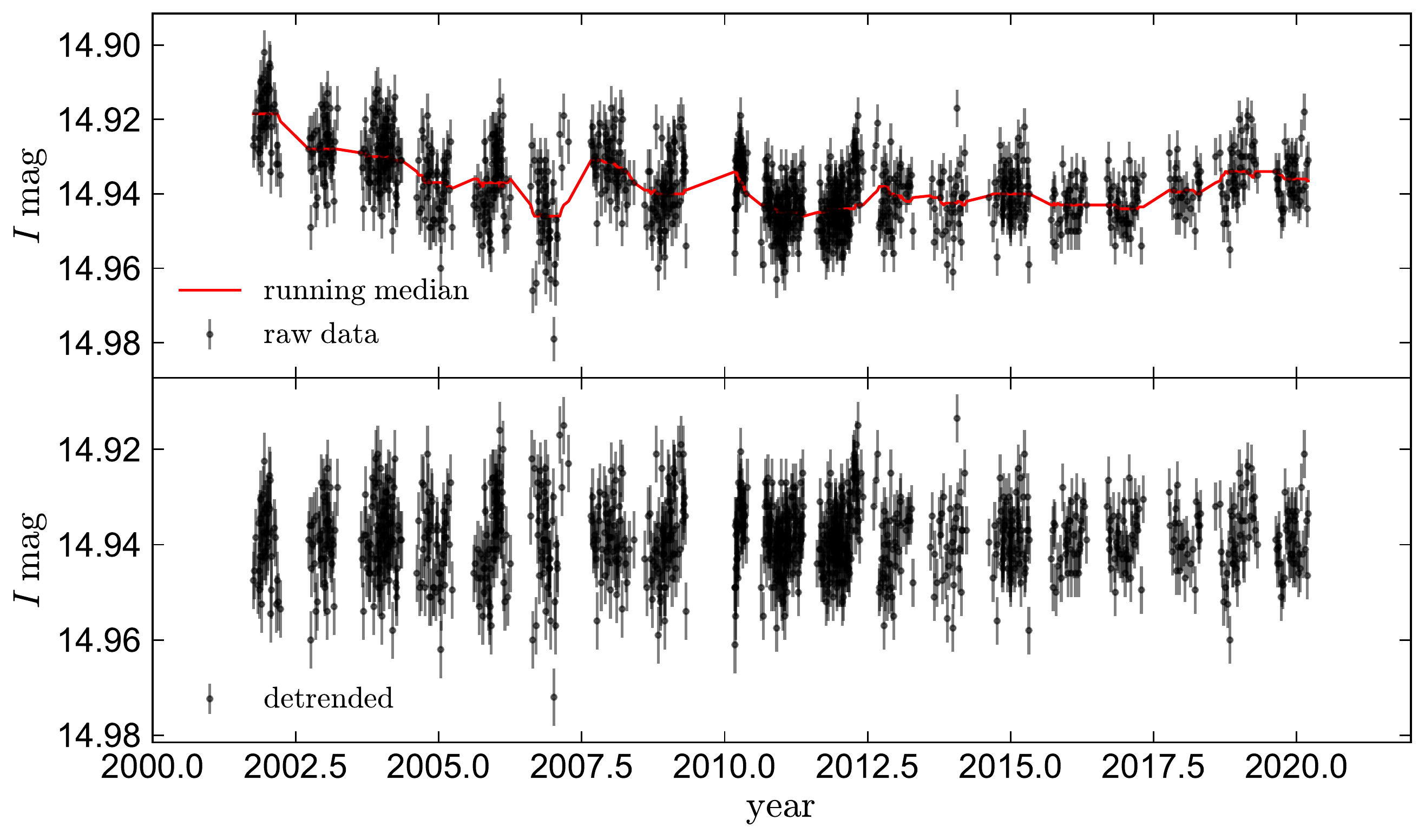}
\caption{{\bf Detrending of the long-term variability}. The variability is observed mostly in the first epoch of observation during 2000-2007 (OGLE III data). Errors represent 1$\sigma$ measurement errors.  The upper panel shows the original light curve, along with the running median filter with a width of 400\,d. The lower panel shows the light curve after detrending.  It is unlikely that the origin of the long-term variation is  intrinsic to the star (see text). } 
\label{fig:detrending}
\end{figure}

The light curve contains evidence of both short- and long-timescale variability. The median $I-$band magnitude varies smoothly on a several-year timescale, with a peak-to-peak amplitude of 0.03 mag. The origin of this long-term variability is uncertain,  though the lack of associated periodicity (e.g., as observed in several Be X-ray binaries \cite{Rajoelimanana2011}), and  the lack of a similar trend in the sparser $V-$band OGLE light curve, suggest it is not intrinsic to the star. In fact, long-term variability is also seen in neighbouring stars in the same OGLE field that are embedded in the nebula, suggesting that it may be related to  nebular contamination.   We remove it in our analysis of the short-timescale variability by detrending the light curve with a running median filter of width 400 days (Supplementary Fig.\,\ref{fig:detrending}).  Given the different timescale of the long-term variation compared to the orbital period, as well as its overall low amplitude, the precise treatment of the long-term variability has no impact on our analysis.

\begin{figure}[!h]
\centering
\includegraphics[width=\textwidth]{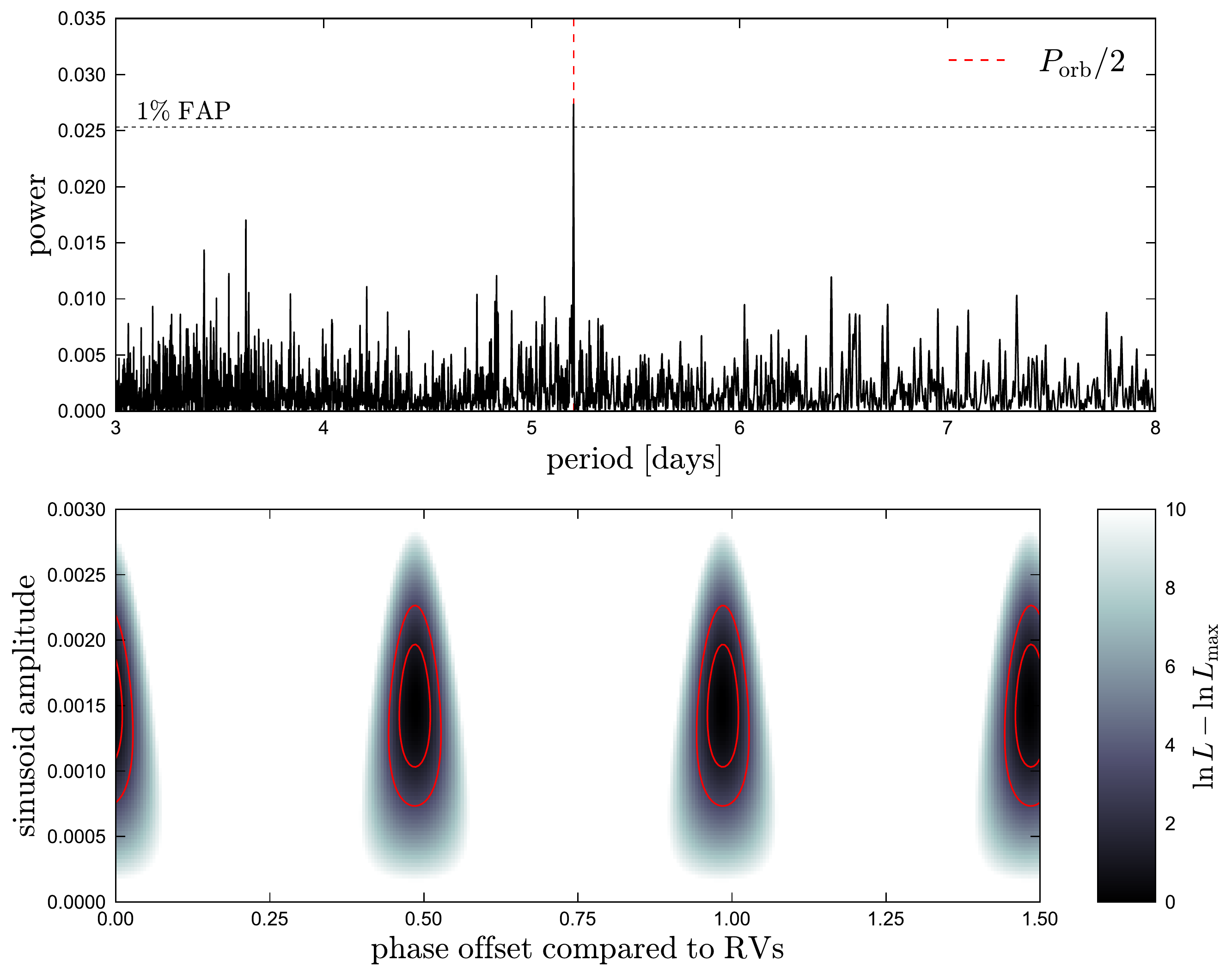}
\caption{{\bf Analysis of the OGLE light curve of VFTS~243.} Top: Lomb-Scargle periodogram of the $I$-band OGLE light curve of VFTS 243. A peak is evident at a period of 5.201 days, exactly half of the orbital period inferred from RVs.  Dashed horizontal line marks a 1\% false alarm probability (FAP). Bottom: results of fitting a sinusoid with $P=5.201$ days to the detrended light curve. Lines enclose 68 and 95\%\ of the integrated 2D probability. The best-fit sinusoid has a phase in excellent agreement with the RV ephemeris (i.e., with peak brightness occurring at the same time as the RV extrema). This suggests the photometric variability is ellipsoidal variability due to tidal distortion of the O star. The marginalized $1\sigma$ constraint on the variability half-amplitude is $A_{\rm ellipsoidal}= 0.0015 \pm 0.0003$. } 
\label{fig:ellipsoidal}
\end{figure}

The top panel of Supplementary Fig.\,\ref{fig:ellipsoidal} shows a Lomb-Scargle periodogram of the $I-$band light curve, which contains a peak at a period of 5.201 days. In the absence of other information, the peak would be only suggestive. However, the fact that it coincides exactly with half the spectroscopic orbital period strongly suggests that it is real and related to the binary's orbital period. The same peak appears in a periodogram of the light curve of LMC517.01.10094, providing further evidence that it traces real variability.

\begin{figure}[!h]
\centering
\includegraphics[width=\textwidth]{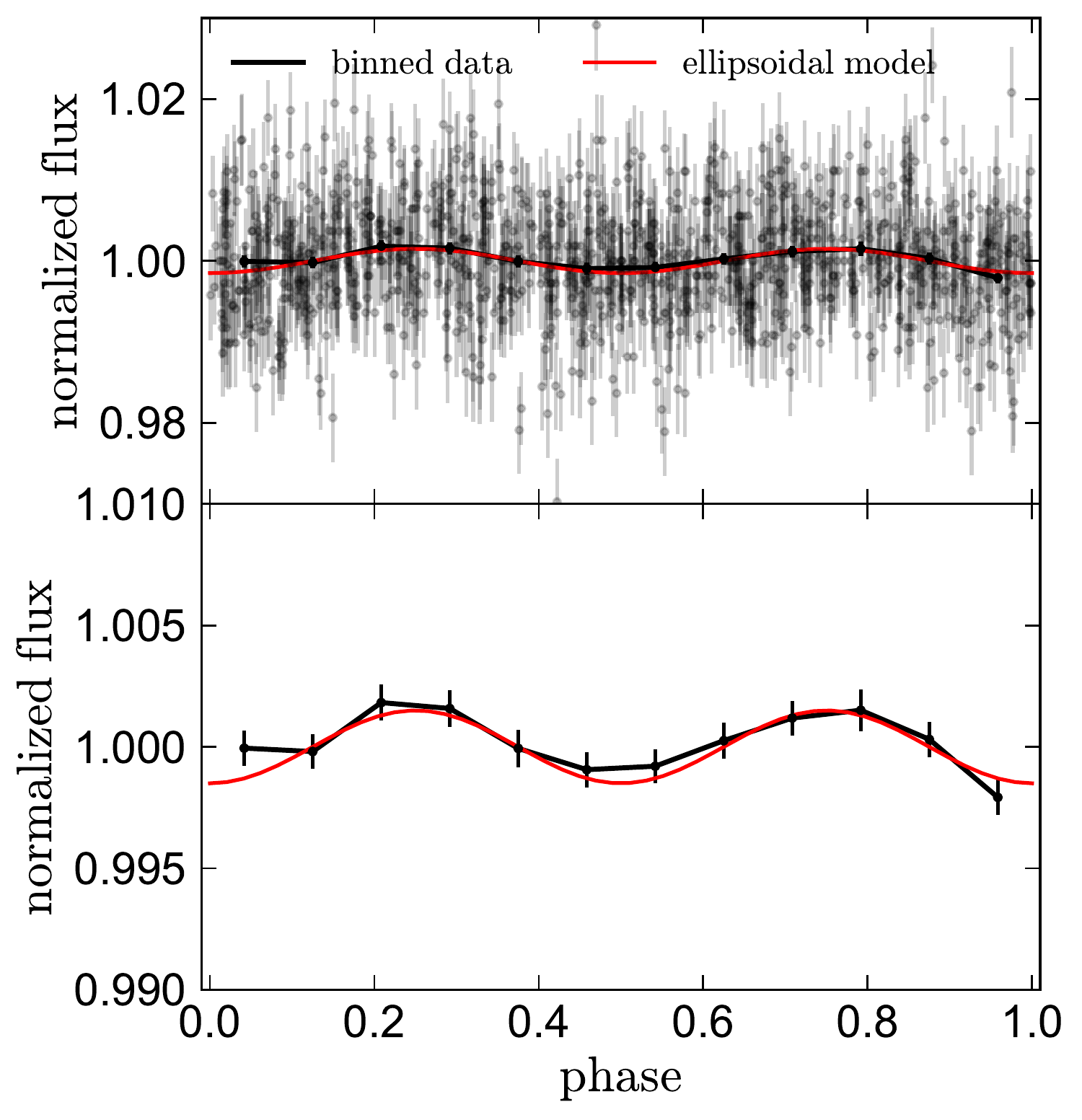}
\caption{{\bf OGLE $I$-band light curve phased to the ephemeris from RVs}.  Errors depict $1\sigma$ measurement errors. The same data are plotted in bins of $\Delta \phi = 0.08$ in the bottom panel (errors are error of the mean).  Red line shows the best-fit sinusoid. The variability amplitude is small compared to the individual photometric errors and is marginally noticeable in the raw data, but is more evident in the binned data.} 
\label{fig:phased_lc}
\end{figure}

We explore this variability further in the bottom panel of Supplementary Fig.\,\ref{fig:ellipsoidal}, where we show results of fitting a sinusoid to the detrended, normalized light curve. We fix the period to 5.201 days and leave the amplitude and phase free. We conservatively inflate the photometric uncertainties to $\approx 0.0075$ mag to yield a reduced $\chi^2$ of 1. The most important conclusion is that the best-fit sinusoid has a phase in good agreement with the RV ephemeris, strengthening the case that the observed variability is due to ellipsoidal variations. In Supplementary Fig.\,\ref{fig:phased_lc}, we show the light curve phased to the ephemeris from the RV analysis. Although the amplitude of the photometric variation is very small, it is evident in both the binned and non-binned light curves.

\begin{figure}[!h]
\centering
\includegraphics[width=\textwidth]{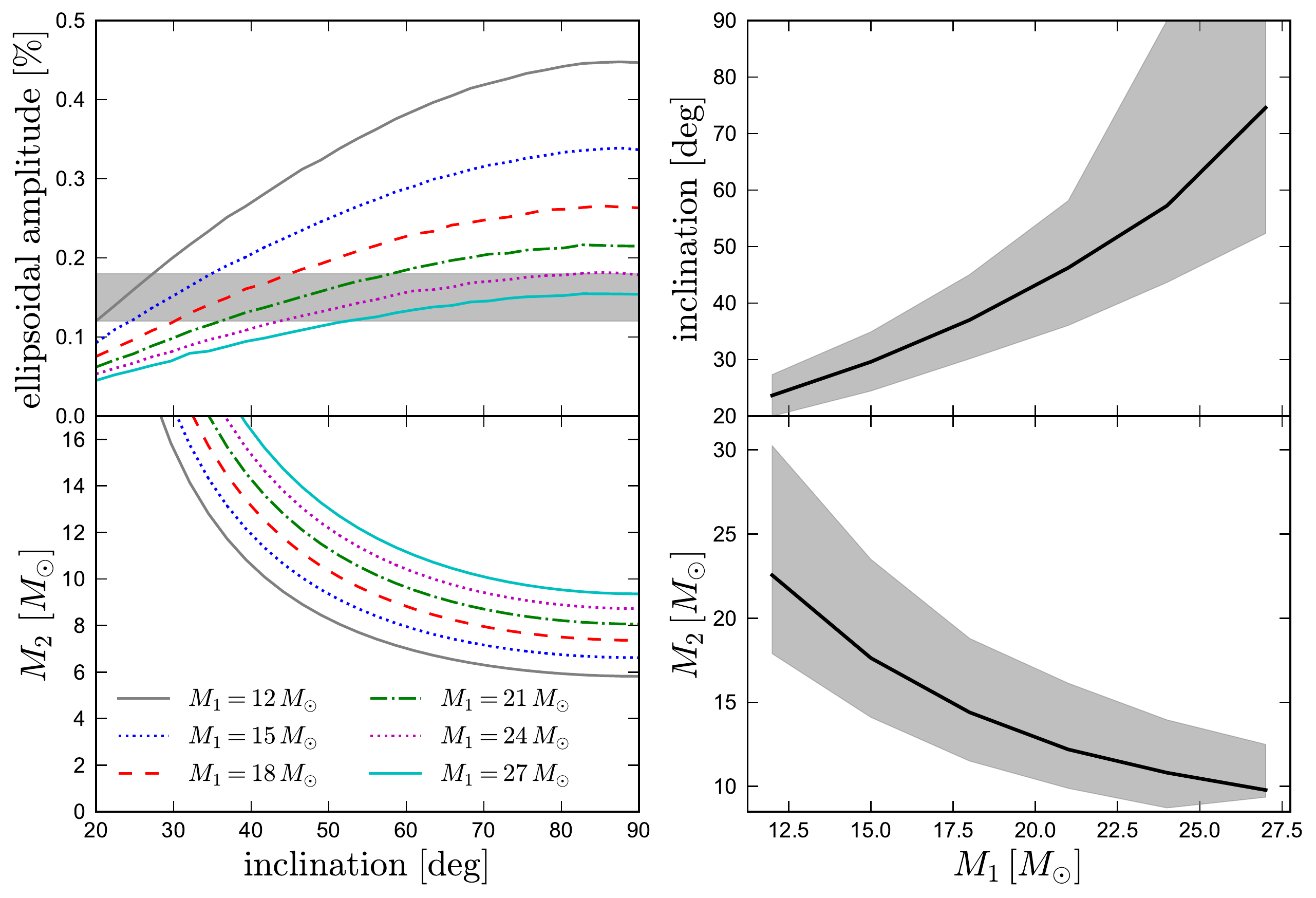}
\caption{{\bf Constraints on component masses and from the observed ellipsoidal variability amplitude.} We assume an O star radius of $10.5 R_{\odot}$. For each value of $M_1$ and orbital inclination, we calculate the companion mass $M_2$ required to reproduce the observed radial velocities (bottom left). For that value of $M_2$, we calculate the predicted ellipsoidal variability amplitude (top left). Gray shading shows the observed value. Right panels indicate the inclination and companion mass needed to match this value for a given $M_1$. Lower values of $M_1$ require a {\it higher} companion mass, because they imply a low inclination. The observed ellipsoidal variability amplitude rules out non-BH scenarios in which the O star is undermassive: a low-mass O star would produce larger-amplitude photometric variability than is observed.  } 
\label{fig:var_amps}
\end{figure}

The ellipsoidal variability amplitude at fixed period depends primarily on the density of the tidally distorted O star (hence $M_1, R_1$) and on the inclination $i$, with weaker dependencies on the mass ratio ($M_2/M_1$) and atmospheric parameters of the star.  At the low amplitude of ellipsoidal variation observed here, the parameters $M_1, M_2$ and $i$ are degenerate, that is, different combinations of $M_1, M_2,$ and $i$ would produce a statistically indistinguishable signal. Hence, instead of fitting the light curve,
we 
explore the range of binary parameters that can match the observed amplitude.
We calculate a set of model light curves of binaries satisfying the orbital constraints that comprise a dark companion and an O star radius of 10.5\,$R_{\odot}$ using the PHOEBE tool \cite{Prsa2005}.  The results are shown in Supplementary Fig.\,\ref{fig:var_amps}.


Supplementary Fig.\,\ref{fig:var_amps} also shows the range of inclination and companion mass that can reproduce the observed variability amplitude and mass function for a given $M_1$.  Evidently, while lower primary masses imply a lower minimum mass on the secondary, they result in an increase of the actual mass of the secondary due to a corresponding decrease in the orbital inclination needed to reproduce the ellipsoidal variability. Hence, based on the light curve, a secondary more massive than $\approx 9\,M_\odot$ is unavoidable. 

We note that the results depend rather sensitively on $R_1$. For example, an increase of $R_1$ within $1\sigma$ ($0.8\,R_\odot$) leads, for a given inclination, to $M_1, M_2$ values that are larger by $\approx 10-15\%$. Similarly, a decrease of $R_1$ yields a decrease of $\approx 10\%$ in $M_1, M_2$. While varying $R_1$ is therefore not negligible quantitatively, it has no bearing on the conclusions of this study, which anyhow do not fundamentally rely on the light curve of VFTS~243.

\section*{X-ray production}

The observed T-ReX Chandra X-ray count rate yields a limit to the observed X-ray luminosity of $\log L_X< 31.84$\,$[{\rm erg}\,{\rm s}^{-1}$].
Typical X-ray attenuation corrections for T-ReX early-type stars are 0.32\,dex (Crowther et al.\ 2022, submitted), such that the intrinsic X-ray luminosity for VFTS~243 is constrained at $\log L_X< 32.16$\,$[{\rm erg}\,{\rm s}^{-1}$]. This amounts to $\log L_X/L < -6.6$, which lies
three orders of magnitude below the candidate high mass X-ray binary VFTS~399, for which $\log L_X/L = -3.4$ \cite{Clark2015}.

Our spectral analysis substantiates the presence of a black hole secondary 
beyond reasonable doubt, and one may therefore wonder whether  the interaction of the stellar wind of the O star with the black hole would  produce X-rays. 
Copious emission is expected for the case of a black hole with a wind-fed accretion 
disk, where the wind matter can be heated to X-ray emitting temperatures. 

Adopting a non-rotating black hole of 9\,$\mso$ as the unseen companion in VFTS\,243,
we calculate a Bondi-Hoyle mass accretion rate of $1.7 \cdot 10^{-11}\,M_{\odot}$ yr$^{-1}$ 
\cite[equation 1]{Bondi1952} when using the unclumped wind mass loss rate and terminal 
wind velocity as given in Table\,1 in the main text.
To gauge whether an accretion disk can form, we follow \cite{Sen2021}. The 
average specific angular momentum $j_{\rm acc}$ of the matter accreted by 
the BH from the wind of the O star is estimated as 
\begin{equation}
j_{\rm acc} = \frac{1}{2} \eta \Omega_{\rm orb} R_{\rm acc}^2 .
\end{equation}
Here, $\eta$ is the efficiency of specific angular momentum accretion, 
for which we use $\eta$=1/3,
$\Omega_{\rm orb}$ 
is the orbital angular velocity, and $R_{\rm acc}$ is the accretion
radius given by equation 8 of \cite{Sen2021}. An accretion disk is expected when the accreted average specific angular momentum exceeds that of a particle in the innermost stable circular orbit (ISCO) around the black hole:
\begin{equation}
j_{\rm ISCO} = \sqrt{G M_{\rm BH} R_{\rm ISCO}},
\end{equation} 
where $R_{\rm ISCO}$ is the radius of the innermost stable orbit. We find 
the specific angular momentum of the accreted wind matter to be $\sim 3/100$ 
times smaller than required for disk formation. Even if the black hole were maximally 
spinning and the efficiency of angular momentum accretion were 100\% 
(see Eqs.\,6 and 7 of \cite{Sen2021}), an accretion disk is 
not expected in VFTS\,243. 

Considering instead radial in-fall of wind matter onto the BH 
leads to optically thin thermal Bremsstrahlung emission escaping 
from the adiabatically heated wind-matter. The expected X-ray 
luminosity in this case is orders of magnitude below that expected 
for disk accretion \cite{Shapiro1986}. While non-ideal processes 
such as turbulence, magnetic fields, and non-radial trajectories 
of accretion may enhance the X-ray emission,
the 
current non-detection of X-rays from VFTS\,243 appears fully 
consistent with the presence of a BH. 

We note that the theoretical expectation regarding the absence of a disk  strongly depends on the wind terminal velocity, which cannot be constrained from existing data. In fact, the presence of accretion disks is not sufficient to ensure the production of copious X-rays. At very low 
accretion rates, accretion disks may become advection-dominated and produce very little X-rays, leading to long periods of quiescence that can last for many years
(e.g., GS~2000+25, \cite{Rodriguez2020}).
While known examples for this partial quiescence are all transient X-ray sources, a population of persistently
X-ray quiet disk-fed BH-binaries has also been proposed \cite{Menou1999}.
Still, VFTS~243 shows no traces of a disk in its optical spectrum, while
the known partly-quiescent BH binaries do so very blatantly  (e.g., \cite{Marsh1994}). 
It therefore  appears likely, both on theoretical and empirical grounds, that no disk is present in VFTS~243. Future ultraviolet observations of VFTS~243 will be crucial to obtain a complete picture regarding the wind and disk properties of VFTS~243.

\section*{Synchronisation, circulation, and rotation}

Tidal interactions between two components of a short-period binary act to synchronise their rotation periods with the orbital period and, on a significantly longer timescale, to circularise the orbit. Given the importance of the small eccentricity to our conclusions, it is essential to consider the impact of circularisation in VFTS~243. The respective timescales (i.e., the e-folding times), $\tau_{\rm sync}$ and $\tau_{\rm circ}$, strongly depend on the orbital separation and internal structure of the stars, as well as on a series of non-trivial assumptions \cite{Zahn1977}. 

Of particular importance in the computation of these timescales is the structure constant $E_2$. The computation of this quantity from a stellar model is not straight forward, but various authors have provided fits to it. In Supplementary Fig.\,\ref{fig:timescales}, we show the synchronization and circularisation timescales computed using a $25~M_\odot$ stellar model computed  with the \texttt{MESA} code \cite{Paxton2011} (Input files to reproduce our \texttt{MESA} calculations are available at \url{doi.org/10.5281/zenodo.6514645}). The simulation is made using the same physical assumptions of \cite{Brott2011} to model LMC stars, except that we ignore mass loss through its evolution to match the current mass estimate. We consider different prescriptions for the structure constant $E_2$, including the one of \cite{Hurley2002}, who computed a fit to the zero age main sequence models of \cite{Zahn1977}, and those of \cite{Yoon2010} and \cite{Qin2018}, who computed fits appropriate for the main sequence evolution of massive stars. Despite the order-of-magnitude variation in the resulting timescales, which reflect the uncertainty on the computation of $E_2$, the circularisation timescale of VFTS 243 is expected to exceed $100$~Myr.

\begin{figure}[!h]
\centering
\includegraphics[width=.8\textwidth]{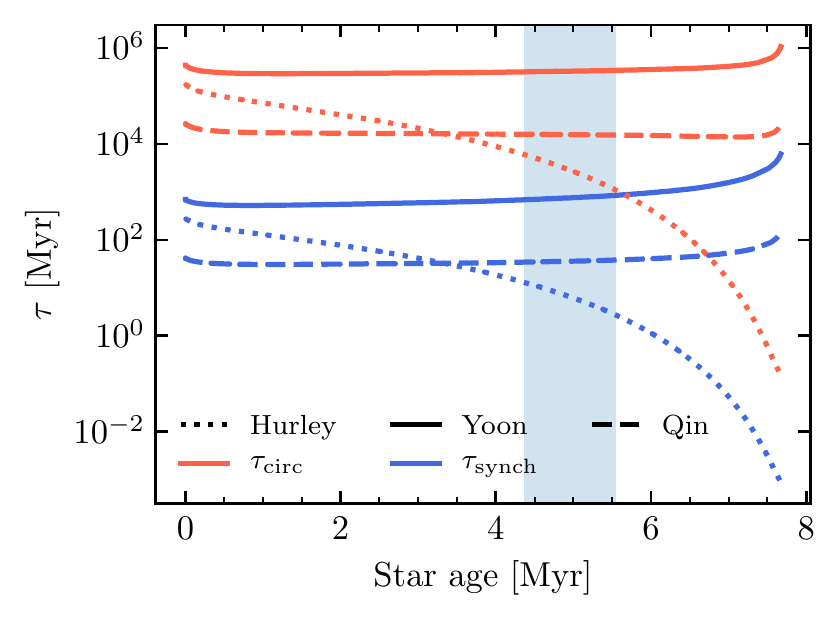}
\caption{{\bf Synchronization and circularisation timescales}.  The computations are performed for a $25M_\odot$ model star during the main sequence, with a fixed orbital period of $10.4031$ days and a companion mass of $10.1~M_\odot$. Timescales are computed using three different fits to the structure constant $E_2$ (see text for details). Vertical band indicates the phase during the evolution where the model has a $T_\mathrm{eff}$ within the observed range of $36\pm 1$ kK.} 
\label{fig:timescales}
\end{figure}

A much more reliable method to assess the significance of tidal circularisation is to consider the ratio of the circularisation and synchronisation timescales, which is independent of $E_2$. Following \cite{Zahn1977}, this ratio can be expressed as:

\begin{equation}
\label{eq:circsync}
    \frac{\tau_{\rm circ}}{\tau_{\rm synch}} = 176\cdot 
    \left(0.1\frac{M_1 R_1^2}{I_1}\right)
    \cdot
    \left(\frac{M_{\rm BH}}{10.1\,M_\odot}\right)
    \cdot
    \left(\frac{M_{\rm tot}}{36.3\,M_\odot}\right)^{-1/3}
    \cdot
    \left(\frac{R_1}{10.3\,R_\odot}\right)^{-2},
\end{equation}
where $I_1$ is the moment of inertia of the primary, defined using a unitless number $k$ as  $I_1 = k\,M_1\,R_1^2$. While $k= 0.1$ is typically assumed for main sequence stars \cite{deMink2013, Hurley2002}, its value can be significantly lower near the end of the main sequence, which would further increase the ratio $\tau_\mathrm{circ}/\tau_\mathrm{sync}$.  Equation\,(\ref{eq:circsync}) is written such that the expressions in parentheses are of the order of unity for VFTS~243, implying that $\tau_{\rm circ}$ is at least 1-2 orders of magnitude longer than $\tau_{\rm sync}$. Hence, if we can argue that the star has not yet synchronised, we may safely conclude that circularisation via tides is negligible.

The derived projected rotation velocity of the primary of $\varv \sin i = 181 \pm 16\,\kms$, combined with our estimates on the orbital inclination from the light curve ($i \approx 40-90^\circ$), implies an equatorial rotation of $\varv_{\rm eq}$ in the range $\approx 180-300\,\kms$. Synchronisation would imply a rotation of $2\,\pi\,R_1\,P^{-1} \approx 50\,\kms$. The system is thus far from being synchronised, following Eq.\,(\ref{eq:circsync}), tidal circularisation can be fully neglected. This means that the near-circular orbit observed today  reflects the orbital eccentricity of the binary immediately after core-collapse. 

It is interesting to note that the rotation of the primary lies above the median of $\langle \varv \sin i \rangle \approx 120\,\kms$ for primary components of binaries analysed in the Tarantula region \cite{Ramirez-Agudelo2015}, consistent with a previous mass-transfer event. However, one may expect mass accretors to reach critical rotation velocities \cite{Packet1981} of  $\approx 500-600\,\kms$. In contrast to the theoretical expectation, post-interaction O-type mass accretors (e.g., in Wolf-Rayet binaries) are commonly observed to rotate at sub-critical velocities, with typical velocities of 200-300\,$\kms$~\cite{Shenar2019, Shara2020}. The reasons for this are yet to be established, but could be related to boosted mass-loss of rapid rotators or magnetic fields produced during mass-transfer \cite{Shara2020}.

\section*{Evolutionary path and final fate}

\begin{figure}[!h]
\centering
\includegraphics[width=\textwidth]{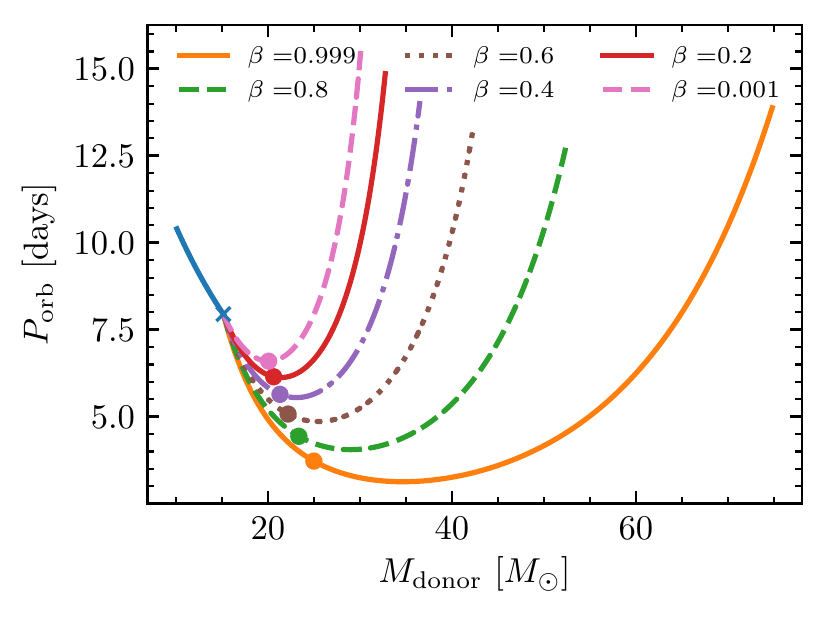}
\caption{{\bf Relationship between orbital period and donor mass for difference mass-transfer efficiencies.} Each line traces the properties of the black-hole progenitor mass ($M_{\rm donor}$) and orbital period $P_{\rm orb}$ for different mass-transfer efficiencies,  from fully conservative ($\beta = 0$) to fully non-conservative ($\beta = 1$). The curves are to be read from left to right, backward tracing the evolution from the current-day values (leftmost point). The curves terminate when the initial mass ratio fulfils $q_{\rm ini} = M_{\rm ini, 2}/M_{\rm ini, 1} < 1/3$, since  binaries with more extreme initial mass ratios are expected to merge.  The black hole progenitor is assumed to have lost 1/3 of its  mass during the preceding Wolf-Rayet phase, which causes the increase of $M_{\rm donor}$ and shrinkage of $P_{\rm orb}$ as the curve advances to the right. Mass transfer terminates at the point marked by a cross (where all curves meet). Filled circles correspond to mass ratios of unity. Mass transfer and the initial configuration could lie anywhere to the right of the filled circles in each respective curves.
} 
\label{fig:evanal}
\end{figure}

It is interesting to consider the possible progenitor masses of the black hole, as well as the initial properties of VFTS~243 and its final fate. One may obtain constraints on the initial parameters by working the evolution of the system backwards analytically for various mass-transfer efficiencies (Supplementary Fig.\,\ref{fig:evanal}). Given the uncertain mass-transfer efficiency, the  progenitor mass could lie anywhere between $\approx 20\,M_\odot$ to  $\approx 80\,M_\odot$, where the smallest values are obtained from conservative mass-transfer, and the highest from non-conservative. We can also conclude that the initial period must have been smaller than $\approx 20\,$d.

To fully solve for the initial properties of this binary, while considering the many uncertainties of binary evolution, one would need more detailed knowledge of the abundances, stellar, and wind parameters, for which high-resolution UV and optical data are required. Here, we provide a potential solution (not necessarily unique), which reproduces the main properties of the binary: $M_2, T_{\rm eff, 1}, \log L_1, P$. We computed a detailed binary evolution model using the Modules for Experiments in Stellar Astrophysics (MESA) stellar evolution code \cite{Paxton2011}. We follow the same set of assumptions as \cite{Brott2011} and \cite{Langer2020}. Given the uncertainties in treatment of rotation and the need for more accurate constraints, we assume non-rotating stars for simplicity and a fixed mass transfer efficiency. Mass transfer efficiency is parameterized in terms of the fraction $\beta$ of mass ejected from the accretor during Roche lobe overflow, carrying its specific orbital angular momentum.

We use the curves provided in Supplementary Fig.\,\ref{fig:evanal} to identify an adequate initial guess on the starting conditions for our simulations (zero age main sequence values of $M_1$, $M_2$, $P_\mathrm{orb}$ and $\beta$). We assume that the mass of the pre-collapse mass donor  at carbon depletion is the final mass of the black hole, as is supported from our findings for little to no ejecta. With this initial guess, we check the properties ($M_2, T_{\rm eff, 1}, \log L_1, P$) of the model at the point of nearest approach in the HR diagram to the observed $T_\mathrm{eff,1}$ and $L_1$ after black hole formation. We iterate on the initial parameters to find a better fitting model by computing numerical partial derivatives of these values with respect to the input parameters, and using a Newton-Raphson solver to determine new initial conditions. We find an adequate solution that reproduces all observables well within $1\sigma$ (but the period, which would require an excessive resolution)  for a mass-transfer efficiency of $1 - \beta = 0.36$ . Our solution has $M_{\rm ini, 1} = 30.1\,M_\odot, M_{\rm ini, 2} = 21.9\,M_\odot$, and $P_{\rm ini} = 3.71\,$d. 

The evolution tracks are shown in a Hertzsprung-Russell diagram (HRD) in  Supplementary Fig.\,\ref{fig:HRD}. The point of closest approach to the observed $T_{\rm eff, 1}, \log L_1$ yields $T_{\rm eff, ev, 1} = 36.5\,$kK, $\log L_{\rm ev, 1} = 5.20\,[L_\odot]$, $M_{\rm BH, ev} = 10.0\,M_\odot$, $P_{\rm ev} = 10.7\,$d, and an age of $7.3\,$Myr. Interestingly, this age is almost twice the age obtained from single-star evolution tracks with BONNSAI (3.9\,Myr). This is a direct result of rejuvenation of the mass accretor with fresh hydrogen imparted from the black-hole progenitor, the original donor. 

\begin{figure}[!h]
\centering
\includegraphics[width=\textwidth]{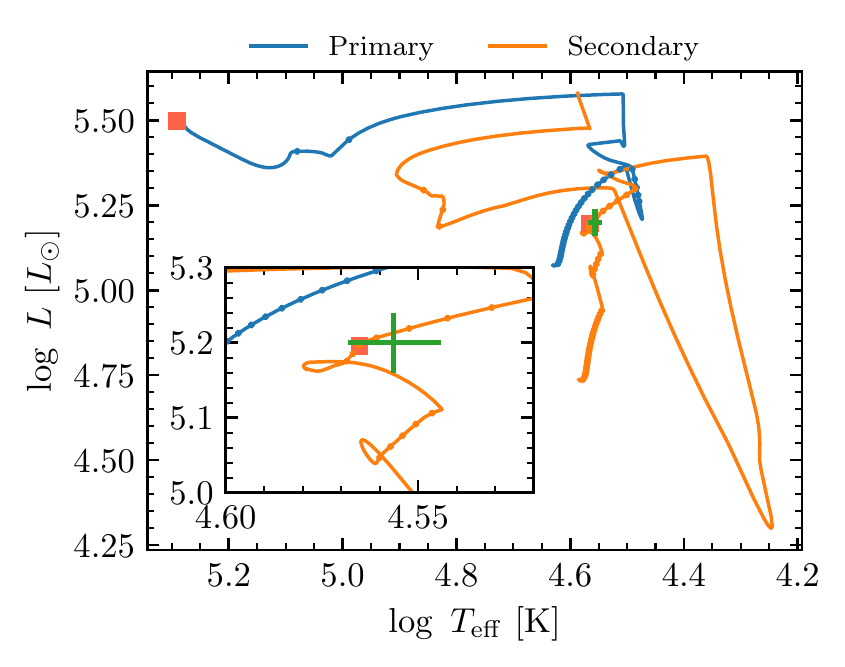}
\caption{{\bf Hertzsprung-Russell diagram (HRD) showing a representative evolutionary scenario for VFTS~243.} Dots indicate steps every $0.2$ Myrs in the evolution.  The binary starts with $M_{\rm ini, 1} = 30.1\,M_\odot, M_{\rm ini, 2} = 21.9\,M_\odot$, and $P_{\rm ini} = 3.71\,$d. A brief phase of rapid case A mass transfer is followed by a longer case A mass transfer phase. The mass transfer comes to a sudden halt as hydrogen is depleted in the core, but initiates quickly again (case AB) as the primary leaves the main sequence and expands. Mass transfer stops when the H-shell burning are removed, and the primary rapidly evolves to the left to become a hot Wolf-Rayet star, until it collapses to form the black hole observed today (denoted by the orange squares), roughly 7.1\,Myr after the formation of the binary. We observe the system today (measurements correspond to green cross) shortly thereafter, at an estimated age of 7.3\,Myr.
} 
\label{fig:HRD}
\end{figure}

We compute our binary model until carbon ignition in the mass accretor (the now O-component). Assuming that the mass accretor would experience direct collapse and no kick (like the mass donor), we find that the binary ends its turbulent life as a BH+BH binary with masses of $10.0~M_\odot+10.0
M_\odot$ and a period of $5.46$ days. Driven by gravitational wave radiation, such a system would merge in $117$\,Gyr, however this depends sensitively upon the masses and final orbital period of the system. If within the observed errors for the current properties of the system the mass ratio has a more extreme value, it is possible that stable mass transfer could harden the orbit sufficiently to result in a binary black hole with a merger delay time smaller than the age of the universe \cite{Marchant2021}.

\section*{Posteriors}\label{sec:posterios}

The posteriors used to compute $M_{\rm min, 2}$,  $M_2$, and $M_{\rm tot}$ are shown in Supplementary Fig.\,\ref{fig:M2minpost}. We note that we do not consider constraints from the light curve when computing the posteriors (Supplementary Fig.\,\ref{fig:var_amps}). However, these constraints are fully consistent with the posteriors derived from spectroscopy alone.

\begin{figure}[!h]
\centering
\includegraphics[width=\textwidth]{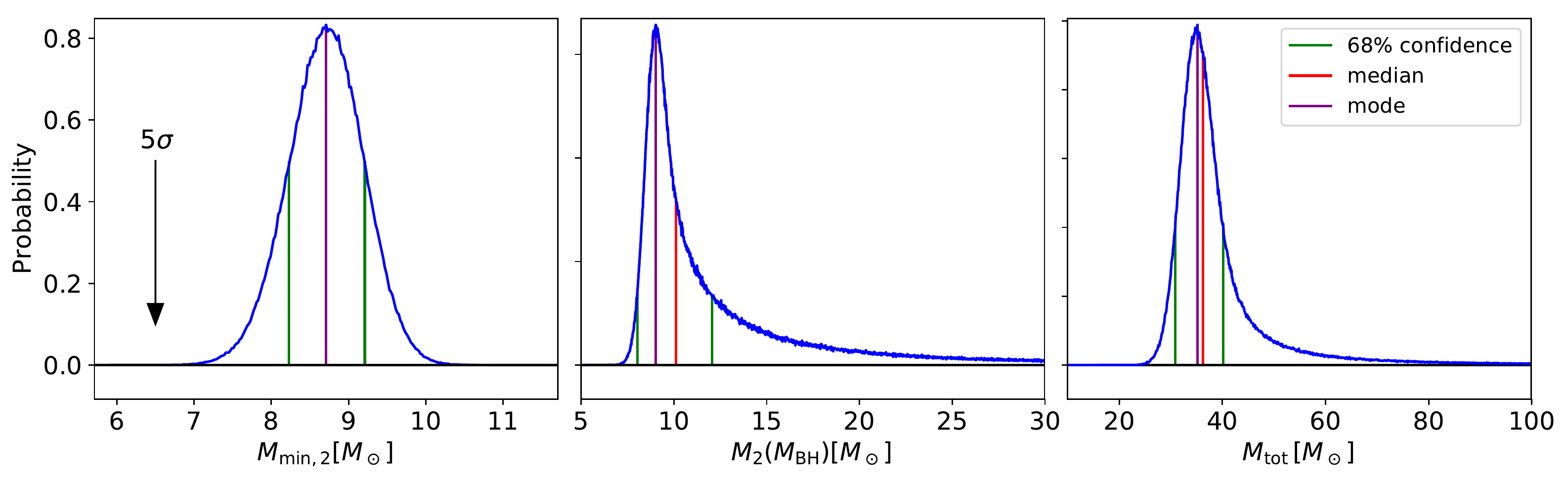}
\caption{{\bf posteriors of the minimum secondary mass $M_{\rm min, 2}$ (left panel), actual secondary mass $M_2$ (middle panel), and total mass of the system $M_{\rm tot}$ (right panel).} The posteriors are computed using the mass function by modelling the distribution of $M_1$ as a Gaussian with a mean of $25.0\,M_\odot$ and a standard deviation of $\sigma = 2.3\,M_\odot$, and assuming a random orientation of the orbital plane.   Marked are the mode (red), median (purple), and 68\% confidence intervals.   } \label{fig:M2minpost}
\end{figure}

\section*{Systematics of spectral analysis}\label{sec:SpecAnDetails}

It is notoriously difficult to derive the masses of stars from spectroscopy, since they depend both on details of data reduction (notably normalisation) as well as on multiple physical mechanisms that operate in a highly non-LTE environment. To explore systematics, we performed thorough comparisons between CMFGEN, PoWR, and the FASTWIND atmospheric model code while analysing the data of VFTS~243. The results for CMFGEN were detailed in Methods. Here, we describe results obtained with the other codes.

\subsection*{PoWR}\label{sec:SpecAnDetails}

\begin{figure}[!h]
\centering
\includegraphics[width=.9\textwidth]{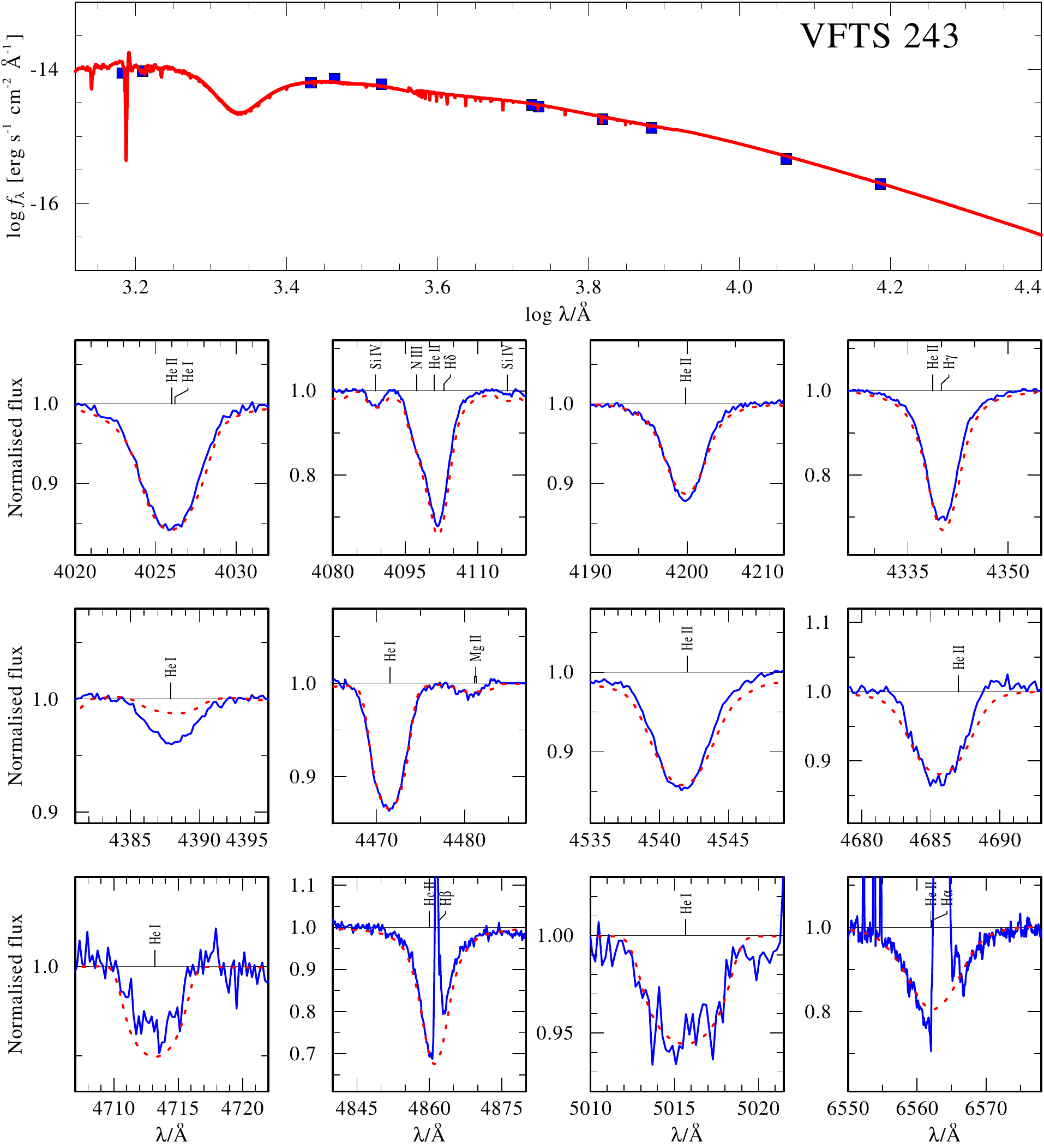}
\caption{{\bf Spectral analysis of VFTS~243 using the PoWR code.} Observations (blue) are compared to best-fitting PoWR model (red), showing the SED in the upper panel, and selected spectral lines in the lower panels.} \label{fig:specanPoWR}
\end{figure}

We used publicly available PoWR spectra \cite{Hamann2003, Sander2015} computed for OB-type stars \cite{Hainich2014} to narrow down the parameter range. The general code assumptions are similar to those described for CMFGEN, with the difference that microturbulence pressure is accounted for.  We  computed several additional models in the region of interest, and show  a representative model (``by-eye fitting") in Supplementary Fig.\,\ref{fig:specanPoWR}. We estimate comparable parameters to those obtained with CMFGEN: $T_{\rm eff} = 35.0\pm1.0\,$kK, $\log g = 3.75\pm 0.10$[cgs], and $\log L = 5.17\pm 0.05$. 
This results in a spectroscopic mass of $22 \pm 5\,M_\odot$, in good agreement with the CMFGEN analysis. The line cores of the Balmer lines are better reproduced for an enhanced (unclumped) mass-loss rate of $ \dot{M} \sqrt{D}= -6.0 \pm 0.2\,[\smy]$. We see clear indications for N enrichment with PoWR as well, with $X_{\rm N} = 0.10 \pm 0.05\%$. A better fit to He\,{\sc i}\,$\lambda 4471$ is reproduced for $\varv_{\rm mic} = 15\,\kms$.

\subsection*{FASTWIND}\label{sec:SpecAnDetails}


As a last test, we also explore the stellar parameters using the non-LTE stellar atmosphere modelling code FASTWIND \cite{Santolaya-Rey1997, Puls2005, Sundqvist2018}. Several simplifications in the non-LTE calculation allow for a much more rapid computation of models (minutes as opposed to hours with CMFGEN and PoWR), which allows for the usage of FASTWIND as an exploratory tool. We employ a fitting algorithm based on genetic evolution (GA), described in detail in  \cite{Abdul-Masih2019, Abdul-Masih2021, Hawcroft2021}. 
We constrain the effective temperature, surface gravity, mass-loss rate, microturbulent velocity and surface abundances of carbon and nitrogen. 
We take a number of the same assumptions as are used in the CMFGEN modelling and optimise the models around the same line list for hydrogen and helium. We also include C\,{\sc iii}\,$\lambda$4069, N\,{\sc iii}\,$\lambda$4379, N\,{\sc iii}\,$\lambda$quad4515, N\,{\sc iii}\,$\lambda$4523, N\,{\sc iii}\,$\lambda$trip4640 and N\,{\sc iv}\,$\lambda$4058. We find best-fit parameters in agreement with those found using CMFGEN, $T_{\rm eff} = 36.5\pm0.5$kK, $\log g = 3.5\pm 0.1$[cgs], log$\dot{M} = -6.2 \pm 0.2$[$M_{\odot}yr^{-1}$], $\epsilon_{N} = 8.2 \pm 0.2$ and $\epsilon_{C} = 7.7 \pm 0.3$. The spectroscopic mass derived using FASTWIND is $13\,M_\odot$, which is at the lower end of our $5\sigma$ uncertainty interval. However, we note that FASTWIND is known to result in $\log g$ values that are systematically 0.12\,dex lower than CMFGEN \cite{Massey2013}. Moreover, the surface gravity  is likely underestimated due to the rotational centrifugal force term and microturbulence pressure (amounting to $\approx 0.1$\,dex together). The results are shown in Supplementary Fig.\,\ref{fig:specanFW1}.

\begin{figure}[!h]
\centering
\includegraphics[width=.75\textwidth]{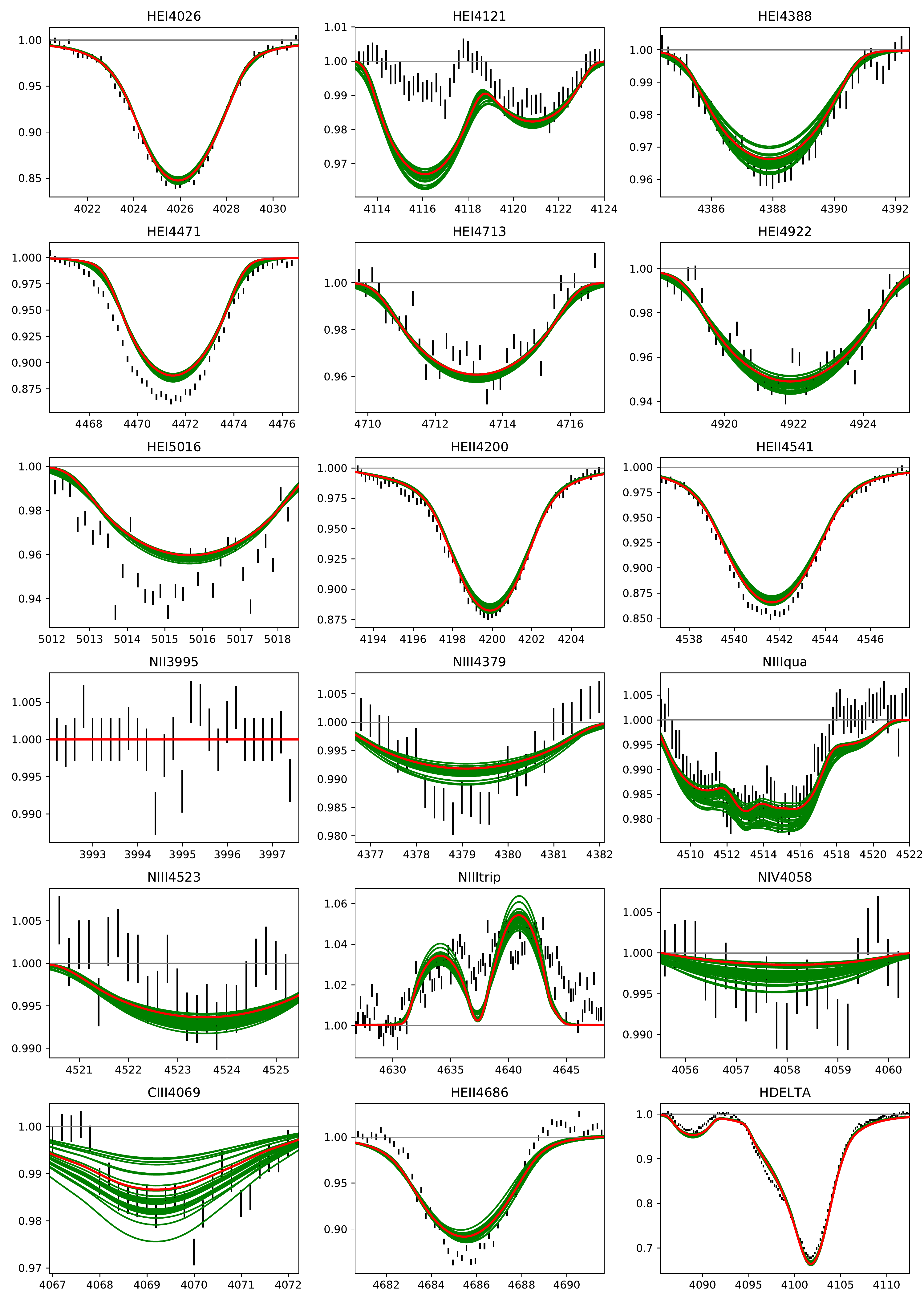}
\caption{{\bf Spectral analysis of VFTS~243 using the FASTWIND code.} Shown are sets of statistically equivalent models (green) and best-fitting model (red) obtained from analysis of VFTS~243 with FASTWIND via genetic algorithms , compared to the data (black data points).  } \label{fig:specanFW1}
\end{figure}

 While the strong nitrogen abundance is confirmed, the carbon abundance is found to be roughly at baseline ($\epsilon_{\rm C} = 7.75$, \cite{Brott2011}), albeit with large errors which can accommodate a factor two decrease. The small discrepancies seen in helium + Balmer lines imply that the star may be slightly enriched in helium, but they could also point towards different microturbulent velocities or wind parameters. Future high-resolution optical and UV data is necessary to robustly derive the abundances and wind parameters of this star.

\bibliography{papers}
\bibliographystyle{naturemag}

\end{document}